%% file: acl_latex.tex
\documentclass[11pt]{article}

\usepackage[preprint]{acl}

\usepackage{times}
\usepackage{latexsym}

\usepackage[T1]{fontenc}

\usepackage[utf8]{inputenc}

\usepackage{microtype}

\usepackage{inconsolata}

\usepackage{graphicx}
\usepackage{hyperref}
\usepackage{url}
\usepackage{array} 
\usepackage[utf8]{inputenc} 
\usepackage[T1]{fontenc}    
\usepackage{hyperref}       
\usepackage{url}            
\usepackage{booktabs}       
\usepackage{amsfonts}       
\usepackage{nicefrac}       
\usepackage{microtype}      
\usepackage{xcolor}         
\usepackage[table]{xcolor}  
\usepackage{amsmath}
\usepackage{xspace}
\usepackage{threeparttable}
\usepackage{graphicx}
\usepackage[most]{tcolorbox}
\usepackage{mdframed}
\usepackage{fancyvrb}      
\usepackage{fvextra} 
\usepackage{makecell}
\usepackage{siunitx}
\usepackage{caption}
\usepackage{multirow}
\usepackage{enumitem}
\usepackage{subcaption}   
\usepackage{booktabs}
\usepackage{longtable}
\usepackage{booktabs}

\usepackage{supertabular}
\usepackage{tabularx}
\usepackage{upgreek}
\usepackage{tocloft}

\newcommand{\agentname}{\texttt{IntrAgent}\xspace}
\newcommand{\agentnamenospace}{\texttt{IntrAgent}}

\newcommand{\benchname}{\texttt{IntraBench}\xspace}
\newcommand{\benchnamenospace}{\texttt{IntraBench}}

\newcommand{\taskname}{\texttt{IntraView}\xspace}
\newcommand{\tasknamenospace}{\texttt{IntraView}}

\title{\agentname: An LLM Agent for Content-Grounded Information Retrieval through Literature Review}

\author{
\textbf{Fengbo Ma\textsuperscript{1,*}},
\textbf{Zixin Rao\textsuperscript{1,*}},
\textbf{Xiaoting Li\textsuperscript{1}},
\textbf{Zhetao Chen\textsuperscript{1}},\\
\textbf{Hongyue Sun\textsuperscript{1}},
\textbf{Yiping Zhao\textsuperscript{1}},
\textbf{Xianyan Chen\textsuperscript{1,$\dagger$}},
\textbf{Zhen Xiang\textsuperscript{1,$\dagger$}}
\\
\\
\textsuperscript{1}University of Georgia, Athens, GA, USA
\\
{\small
\textsuperscript{*}Equal contribution;
\textsuperscript{$\dagger$}Corresponding authors: 
\href{mailto:xychen@uga.edu}{xychen@uga.edu}, 
\href{mailto:zxiangaa@uga.edu}{zxiangaa@uga.edu}
}
}

\begin{document}

\maketitle

\begingroup
\renewcommand\thefootnote{}
\footnotetext{\agentname Project page: \url{https://intragent.github.io/}. \agentname Code: \url{https://github.com/FengboMa/IntrAgent}. \benchname Dataset: \url{https://huggingface.co/datasets/IntrAgent/IntraBench}.}
\endgroup

\begin{abstract}
Scientific research relies on accurate information retrieval from literature to support analytical decisions.
In this work, we introduce a new task, \textit{INformation reTRieval through literAture reVIEW} (\taskname), which aims to automate fine-grained information retrieval \textit{faithfully} grounded in the provided content in response to research-driven queries, and propose \agentname, an LLM-based agent that addresses this challenging task.
In particular, \agentname is designed to mimic human behaviors when reading literature for information retrieval -- identifying relevant sections and then iteratively extracting key details to refine the retrieved information.
It follows a two-stage pipeline: a \textit{Section Ranking} stage that prioritizes relevant literature sections through structural-knowledge-enabled reasoning, and an \textit{Iterative Reading} stage that continuously extracts details and synthesizes them into concise, contextually grounded answers.
To support rigorous evaluation, we introduce \benchnamenospace, a new benchmark consisting of 315 test instances built from expert-authored questions paired with literature spanning \textit{five} STEM domains.
Across seven backbone LLMs, \agentname achieves on average 13.2\% higher cross-domain accuracy than state-of-the-art RAG and research-agent baselines.

\end{abstract}

\section{Introduction}
In modern scientific research, accurately extracting metadata, experimental setups, and contextual knowledge from prior literature is essential.

However, such information retrieval is challenging due to the complexity of scientific literature~\citep{huang_survey_2025}, which often requires deep domain expertise and substantial time to read and interpret~\citep{wu_retrieval-augmented_2025,lu_efficient_2024,li_enhancing_2024, wang_crafting_2024}.
Therefore, an automated system for researchers to accurately and efficiently extract relevant information from literature can potentially transform research practices across many scientific disciplines.

Motivated by this need, we propose a new task called INformation reTRieval through literAture reVIEW (\taskname) -- given a literature and an information retrieval query, extract and synthesize information \textbf{faithfully grounded in the provided content}.
While \taskname follows the general format of Content Question Answering (CQA)~\cite{jin-etal-2019-pubmedqa,mihaylov_can_2018,welbl_crowdsourcing_2017,chen_hybridqa_2021,reddy_coqa_2019,chen_reading_2017}, it introduces greater challenges due to the structural complexity and domain-specific language of scientific literature, in contrast to the more generic knowledge chunks typically used for standard CQA~\cite{dasigi_dataset_2021,lu2022learnexplainmultimodalreasoning,joshi_triviaqa_2017,he_olympiadbench_2024,rajpurkar_know_2018,rein_gpqa_2023,jin2021disease,kritharaBioASQQAManuallyCurated2023a}.
Moreover, in practical applications of \tasknamenospace, it is crucial to avoid hallucination by explicitly acknowledging when the requested information is not present in the literature.
Due to these critical differences, existing methods for conventional CQA tasks are insufficient for addressing the new challenges of \tasknamenospace~\cite{wu_mathchat_2024,lu_mathvista_2024,wang_mathcoder_2023}.
\textit{Direct LLM querying} for specific information retrieval -- even with models fine-tuned for scientific domains~\cite{zhang-etal-2024-comprehensive-survey} -- struggles to manage the overwhelming information contained in a full literature.
\textit{Retrieval-augmented generation} (RAG) approaches attempt to address this limitation by selecting a small set of text chunks deemed most relevant to the query~\cite{10.5555/3495724.3496517,wang-etal-2024-rag,ye_r2ag_2024}, typically based on semantic similarity, and then using these chunks to prompt the LLM.
However, such methods rely solely on surface-level semantic similarity, often failing to align domain-specific terminology in the query with the actual relevant content, and ignore the rich structural organization inherent to scientific documents.
Recently, numerous LLM-based agents have been developed to tackle complex tasks~\cite{m_bran_augmenting_2024,yinAgentLumosUnified2024,wu-etal-2024-sparkra,tangEfficientInterpretableCompressive2023,nguyen2025teamworkmakesdreamwork}, including those involving scientific literature, such as literature search, summarization, research idea exploration, and academic writing assistance.
Compared to standalone LLMs, these agents leverage the reasoning capabilities of LLMs for comprehensive task planning and interact with external tools and environments to enhance task execution~\cite{whitfield_elicit_2023}.
However, the scientific QA task addressed by these agents \textit{fundamentally differs} from our \taskname. 
They aim to answer general scientific questions through external resource exploration, whereas \taskname focuses on retrieving and reasoning strictly constrained to a provided paper. 
Thus, these methods perform poorly on \taskname, as will be demonstrated in our experiments.

To close this gap, we propose an LLM agent, \agentnamenospace, as the first specialized solution to \tasknamenospace.
The key idea behind \agentname is to emulate human behavior when reading technical literature for information retrieval -- identifying the most relevant sections and progressively accumulating details until the query is fully addressed \cite{effectiveness-chunking,miller1956magical}.
Accordingly, \agentname comprises two main stages. First, in the \textit{section ranking} stage, paper sections are reordered based on structure-aware reasoning by the LLM to reflect their relevance to the query.
Second, in the \textit{iterative reading} stage, the reordered sections are sequentially accessed to extract relevant details, continuing until the accumulated information is deemed, via LLM reasoning, sufficient to answer the query.
Notably, \agentname incorporates several novel designs to address the unique characteristics and challenges of \tasknamenospace.
For example, we introduce a hierarchy preservation step that captures the structural organization of the research literature, enabling more comprehensive reasoning during section ranking.
Additionally, during iterative reading, we design a sufficiency check mechanism to assess whether the extracted details are adequate for answering the query, ensuring the faithfulness by reducing the risk of hallucination.

In addition, we introduce \benchnamenospace, the first benchmark designed to evaluate \taskname approaches, including our proposed \agentnamenospace. 
The benchmark consists of 315 test instances drawn from five scientifically and societally significant domains—physics, earth science, public health, engineering, and material science. 
We further assess \agentnamenospace against state-of-the-art RAG systems and literature-oriented agents, demonstrating its superior performance across all domains. 
Our key contributions are summarized below:

\begin{itemize}[leftmargin=*]
\setlength\itemsep{0em}
\item We introduce \tasknamenospace, a novel task for accurate, automated, and \textit{content-grounded} information retrieval from a provided scientific literature, 

and propose \agentnamenospace, an LLM agent specifically designed to tackle this task.
\item We develop a novel two-stage pipeline for \agentnamenospace, consisting of section ranking and iterative reading, designed to mimic human reading behavior.
We introduce a hierarchy preservation mechanism to help the agent leverage structural knowledge for more effective section ranking, and implement a sufficiency check to mitigate hallucination during iterative reading.
\item We propose \benchnamenospace, the first benchmark for evaluating \tasknamenospace, consisting of 315 test instances across five impactful domains. 
\item We evaluate \agentname on \benchname and show that it outperforms both state-of-the-art RAG and literature-agent baselines across representative backbone LLMs in average cross-domain accuracy.
\end{itemize}

\section{Related Work}

\paragraph{Retrieval-Augmented Generation (RAG)} 
RAG enhances LLMs by retrieving relevant documents for response generation~\cite{gao_retrieval-augmented_2024}, enabling external knowledge integration in knowledge-intensive tasks~\cite{wu_retrieval-augmented_2025}.
The vanilla RAG framework adopts a three-stage pipeline consisting of indexing, retrieval, and generation~\cite{10.5555/3495724.3496517}, but it faces challenges such as retrieval noise, hallucinated outputs, and limited reasoning over retrieved content~\cite{zhu_halueval-wild_2024}.
Recent RAG variants improve embedding quality, retrieval accuracy, and context control through techniques such as re-ranking~\cite{ye_r2ag_2024}, dynamic embeddings~\cite{jiang2024longragenhancingretrieval}, contextual retrieval~\cite{anthropic2024contextual}, external memory~\cite{li_enhancing_2024, wang_crafting_2024}, and modular pipelines~\cite{lu_efficient_2024}.
However, most of these approaches still follow a flat retrieve-then-generate architecture, which limits their effectiveness for fine-grained information retrieval from structured scientific content, as required by \tasknamenospace.

\paragraph{LLM Agent for Literature Tasks}

Many domain-specific research agents have been developed across scientific disciplines, including chemistry~\cite{m_bran_augmenting_2024,tang_chemagent_2025}, engineering~\cite{zhang_honeycomb_2024,10943742}, mathematics~\cite{wu_mathchat_2024}, and bioinformatics~\cite{xin_bioinformatics_2024}; but they are tailored to tasks within a single field.
There also exist agents for cross-domain literature-related tasks other than \taskname, such as literature search~\cite{agarwal2024litllm}, research ideation~\cite{lu_ai_2024}, and end-to-end scientific writing~\cite{schmidgall2025agentlaboratoryusingllm,shao2024assistingwritingwikipedialikearticles}. 
However, most of these agents are not directly applicable to \taskname, which presents distinct goals and requirements. 

While some agents for general scientific QA via literature search, such as PaperQA~\cite{lala_paperqa_2023}, PaperQA2~\cite{PaperQA2}, QASA~\cite{lee2023qasa}, and SciMaster~\cite{chai2025scimaster}, can be adapted to \taskname by disabling their retrieval component and supplying the same paper as input, these systems are not designed for \taskname and therefore cannot match the performance of our \agentnamenospace, as will be demonstrated in our extensive evaluation.

\section{Proposed \taskname Task}\label{sec:intraview}

Information retrieval from scientific literature is critical to a wide range of subjects.
The retrieved information could be used to guide the experimental design, hypothesis refinement, simulation configurations, statistical methodology, results validation, and other downstream decision-making in research workflows~\cite{rothstein_ultra-sensitive_2024,kumar_precision_2024,yadav_glad_2022,senapati_affordable_2024}.

In this work, we propose a novel task, INformation reTRieval through literAture reVIEW (\taskname), which aims to accurately extract key information from a provided research paper in response to a specific query.
For example, given a Surface-enhanced Raman spectroscopy (SERS) paper with query \textit{``What is the excitation laser wavelength used for SERS measurements?''} one should respond with a specific laser wavelength \textit{solely} based on the experimental context described in the paper~\cite{zhao_unveiling_2024}.
Formally, we construct \taskname as a CQA problem~\cite{jin-etal-2019-pubmedqa}.
The objective is to build an automated system that, when given a literature $C$ and a research-driven question $Q$, generates an accurate and \textit{content-grounded} answer $A$ by identifying the most relevant information in $C$ without hallucination. 

Compared to \textbf{existing CQA tasks}, \taskname differs in two aspects: (a) it provides the full literature rather than a pre-selected or processed chunk, with the relevant information potentially appearing anywhere in the literature or \textit{not at all}; and (b) it tackles domain-specific queries that may require cross-referencing multiple sections beyond where the final answer resides.
Compared with other literature-related tasks, such as \textbf{scientific QA via literature search}, \taskname{} emphasizes \textbf{faithful} information retrieval restricted to the provided content, independent of the external validity of its scientific claims.
In contrast to other existing tasks such as PeerQA\cite{baumgartner-etal-2025-peerqa}, where answers may involve author intent, argumentative reasoning, or external domain knowledge beyond the paper, \taskname strictly requires that all answers be directly grounded in the provided content, necessitating structured, multi-stage reading rather than generative reconstruction.
Therefore, existing approaches for those tasks cannot effectively tackle \taskname, as will be shown by our extensive evaluation.

\section{Proposed \agentname Framework}\label{sec:intragent}

\subsection{Overview of \agentname}\label{subsec:method_overview}

As the first agent framework specially designed for \tasknamenospace, \agentname employs ``\textit{mindset bionics}'' approach to emulate the natural reading workflow of humans during information retrieval -- it begins with a guiding question, infers the section most likely to contain the answer, extracts the key information, evaluates whether the question has been sufficiently addressed, and iterates through additional sections as needed~\cite{miller1956magical,effectiveness-chunking}.

\begin{figure*}[t!]
  \centering
  \includegraphics[width=\linewidth]{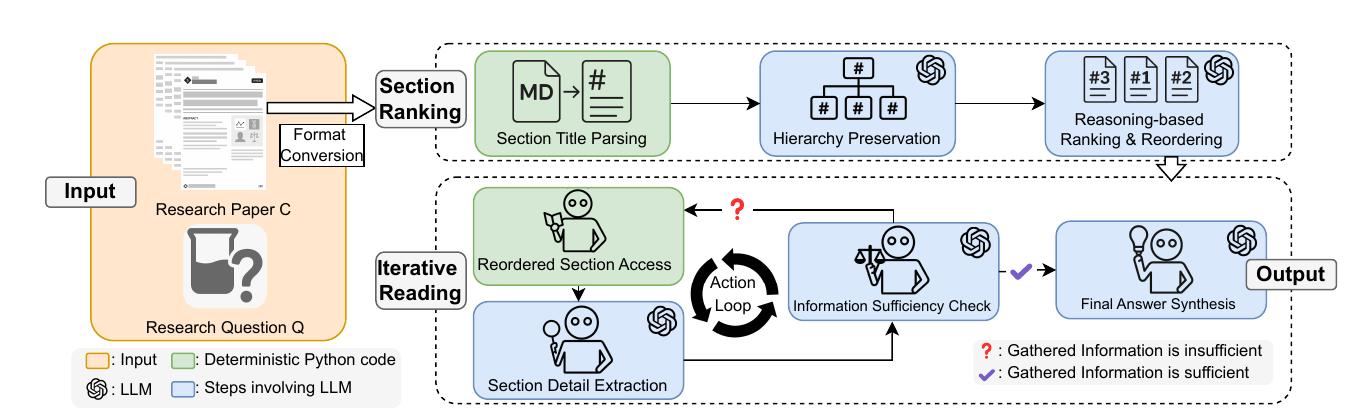}
  \caption{Overview of the \agentname pipeline containing two stages: \emph{Section Ranking} (top) reorders the paper’s sections by relevance to the Research Question $Q$, while \emph{Iterative Reading} (bottom) steps through ranked sections, extracting information until gathered information is sufficient.
  }
  \label{fig:intraview-pipeline}
\end{figure*}

Accordingly, \agentname comprises two major stages -- \textit{section ranking} (Section ~\ref{subsec:Section Ranking}) and \textit{iterative reading} (Section ~\ref{subsec:Iterative Reading}) -- as illustrated in Figure~\ref{fig:intraview-pipeline}.
In the section ranking stage, the agent identifies the most relevant sections of a literature using an LLM by structure-aware reasoning.
In the iterative reading stage, it repeatedly gathers information and evaluates sufficiency until the input query can be adequately answered.

Our design of \agentname offers several advantages over approaches for other CQA tasks, such as RAG~\cite{10.5555/3495724.3496517}, making it well-suited for \taskname:
1) Effective context prioritizing: Unlike RAG’s semantic-similarity-based chunk ranking, our reasoning-based section ranking more precisely locates the details relevant to the query within the literature.
2) Explicit hallucination mitigation: The sufficiency check in the iterative reading stage determines whether additional reading is necessary, thereby explicitly reducing hallucination by ensuring that only substantiated answers are returned.
A toy example of \agentname handling a SERS-related query is shown in Figure~\ref{fig2}.

\subsection{Section Ranking}\label{subsec:Section Ranking}

Scientific research literature typically follow a well-defined \textit{section hierarchy}: parent-level headings convey broader topics, while sub-level headings provide more specific details.
Unlike flat, semantic-similarity-based RAG that overlooks document structure and often fails to align a scientific question with the relevant sections~\cite{10.5555/3495724.3496517}, \agentname leverages this structural knowledge for reasoning-based ranking, prioritizing sections relevant to the question through three key steps:

\textbf{Section Heading Parsing}
To standardize structural information across literature with diverse templates, we convert each input literature $C$ into a Markdown-formatted version $C'$ with systematic markers for section headings.
For papers provided as PDFs, which is the most common case, we use minerU~\cite{wang2024mineruopensourcesolutionprecise}, a visual model for layout and section detection, for the conversion.

\textbf{Hierarchy Preservation}
This step aims to construct a \textit{section tree} that represents the hierarchical structure of the literature.
This tree will support: 1) LLM-based reasoning for section ranking, and 2) structured text parsing for iterative reading.
We begin by extracting all section and subsection headings from the Markdown-parsed text $C'$ to form an initial heading set $\mathcal{H}_0$\footnote{Markdown files from minerU uniformly prepend a single pound sign to all section and subsection titles.}.
Next, we prompt an LLM to infer the hierarchical relationships among these headings, treating each parent section as a node with child nodes corresponding to its subsections.
The prompt is detailed in Appendix~\ref{prompt:Hierarchy-Preservation-Prompt}.
The LLM returns a set of paths from the root to each node in the hierarchy.
To eliminate redundant nodes, we remove paths where a parent section is immediately followed by a subsection without intervening content.

The resulting filtered set of headings is denoted as $\mathcal{H} = \{h_1, h_2, \dots, h_n\}$.

\textbf{Reasoning-Based Ranking} 

Given the filtered heading set $\mathcal{H}$ and the research question $Q$, the model is prompted to reason about which section is most likely to contain the information needed to answer $Q$ -- this design mirrors natural human reading behavior (see Appendix~\ref{prompt:Reasoning-Based Ranking-Prompt} for the detailed prompt).

The output of this step is a ranked list of indices \( R = [r_1, \dots, r_n] \).
Using $R$, we reorder the sections to construct a list of (heading, text) pairs $C_R = [(h_{r_1}, t_{r_1}), \dots, (h_{r_n}, t_{r_n})]$, where $t_{r_i}$ denotes the text in the literature corresponding to section $h_{r_i}$.
These reordered sections will be sequentially accessed during the iterative reading stage described next.

\subsection {Iterative Reading}\label{subsec:Iterative Reading}

At each step, the agent selects its next action from a predefined set: \textit{reordered section access}, \textit{section detail extraction}, and \textit{information sufficiency check}, based on reasoning over the outcomes of the previous step.
This action space is carefully designed to enable targeted retrieval of key information relevant to the input query while suppressing hallucination.

\textbf{Reordered Section Access} When reading is initiated or a section \( (h_{r_i}, t_{r_i}) \) has been fully processed, the agent retrieves the next section \( (h_{r_{i+1}}, t_{r_{i+1}}) \) from the list $C_R$.
This ensures that sections are read in descending order of estimated relevance to the research question \( Q \), maintaining a focused and efficient reading trajectory.

\textbf{Section Detail Extraction} 

The agent examines the current section \( (h_{r_i}, t_{r_i}) \) to extract information relevant to \( Q \). This action is designed to identify and record key scientific details \( D_i \), including terminology, numerical data, experimental results, measurements, statistical indicators, conclusions, and any comparative or causal statements that explicitly address the research question. Each detail in \( D_i \) is anchored to its original sentence and stored in a short-term memory for final answer synthesis. See Appendix~\ref{prompt:Section-Detail-Extraction-Prompt} for detailed prompts.

\textbf{Information Sufficiency Check} 
This step governs the action loop of iterative reading by determining whether more sections need to be accessed.
The agent uses an LLM to reason over the details \( D_i \) gathered in the current iteration and assess whether they are sufficient to answer the research question \( Q \).
The reading loop continues with the next section if the LLM outputs \texttt{NO}; otherwise, it terminates if \texttt{YES} is returned.
Notably, our prompt here includes explicit instructions to avoid speculation and hallucination, as detailed in Appendix~\ref{prompt:Information-Sufficiency-Check-clvl2-Prompt}.
The relevant evidence for a given query may be distributed across multiple non-adjacent sections of a paper. The sufficiency check explicitly prevents premature termination by requiring the agent to continue reading until sufficient supporting information has been accumulated across sections, enabling reliable \textit{cross-section} synthesis.
Moreover, we introduce a confidence-based mechanism that supports three reading styles: \textit{conservative}, \textit{balanced} (default), and \textit{aggressive}, allowing the user to control the operational overhead -- conservative reading accesses fewer sections, while aggressive reading explores more.
Relevant prompts are detailed in Appendix~\ref{prompt:Confidence-Level}.

Once the sufficiency check terminates the action loop after reading $m$ sections, the agent invokes an LLM to synthesize the final answer \( A = \text{LLM}(D_1, \dots, D_m, Q) \) using the accumulated details \( \{D_1, \dots, D_m\} \) during the iterative reading.
Detailed prompts are provided in Appendix~\ref{prompt:Final-Answer-Prompts}.

\section{Proposed \benchname for Evaluation}\label{sec:intrabench}

\subsection{Overview of \benchname}

As a novel task, \taskname lacks dedicated evaluation benchmarks.
Thus, we propose \benchname, a benchmark specially designed for \taskname, which demands expert-level understanding of domain-specific contexts.
\benchname consists of 315 test instances, derived from expert-curated insights and grounded theory~\cite{hallberg2010literature} -- paired with research literature from five highly impactful domains spanning physics, earth science, public health, engineering, and material science.
Although focusing on different tasks, \benchname contains more test instances from broader scientific fields compared with existing benchmarks for QA via literature search (see Table \ref{tab:benchmark_comparison}).
More details on benchmark construction, including the domain choices, are deferred to Appendix~\ref{sec: Details about the dataset}.

\begin{table}[t]
\centering
\small
\begin{tabular}{lccc}
\hline
\textbf{Benchmark} & \textbf{\# Instances}  & \textbf{Domains}\\
\hline
LitQA & 50  & Bio  \\
LitQA2 & 248  & Bio \\
\textbf{\benchname} & \textbf{315} & Phys, ES, PH, Engr, MS \\
\hline
\end{tabular}
\caption{Comparing \benchname with existing benchmarks for scientific QA via literature search (Phys: Physics, ES: Earth Science, PH: Public Health, Engr: Engineering, MS: Material Science, Bio: Biology).}
\label{tab:benchmark_comparison}
\end{table}

\subsection{LLM-Grounded Multiple-Choice Evaluation for \benchname}\label{subsec:bench_evaluation}

In real-world applications, \taskname is expected to produce concise, \textit{free-form} answers rather than selecting from predefined choices.
However, this presents two major challenges for evaluation.
1) Scientific terminologies often involve abbreviations and synonyms -- for example, terms like AgNP, silver nanoparticle, and silver nanorod may all refer to the same surface-enhanced substrate treatment in a SERS experiment. 
2) When retrieved information is numerical or factual, precise accuracy is critical.
Traditional string evaluation methods, such as BLEU~\cite{papineni-etal-2002-bleu} or ROUGE~\cite{lin-2004-rouge}, are unsuitable since they rely on surface-level text similarity and often fail to capture semantic correctness in specialized scientific contexts.

To address these challenges, we adopt a multiple-choice format when creating questions for \benchnamenospace, though the evaluated methods must still generate short answers without being shown the multiple-choice options.
During evaluation, an LLM maps each generated short answer to the most relevant multiple-choice candidate.
This relevance-based mapping effectively addresses the first challenge of varied terminologies such as abbreviations, synonyms, or domain-specific expressions referring to the same concept.
If the LLM cannot confidently align the generated answer with any provided choice (due to insufficient relevance), we use a fallback label -- ``None of the above'' -- to avoid misclassification of ambiguous responses.

The final evaluation metric is accuracy based on the multiple-choice mapping. Empirically, this method demonstrates strong alignment (Average correctness agreement of 63/65 on the Physics dataset using GPT-4.1) with manual mappings by domain experts; see Section~\ref{subsec:OOD} for details.

\subsection{Construction of \benchname Dataset}

\paragraph{Selection of Papers}
For each domain, papers are manually selected by a corresponding expert on our team.
To reduce selection bias, each expert first curates a larger pool of familiar papers, from which five are randomly chosen.
The selected papers are constrained to impactful, peer-reviewed journals to ensure their authority.
Experts’ familiarity with these papers ensures accurate annotation of the ground-truth answers for the associated questions.

\paragraph{Creation of Questions}
We aim to capture both technical depth and conceptual complexity through expert-level inquiry.
To this end, the questions are generated by our domain experts based on their natural reading practices, i.e., what information they would seek from a paper within their field.
These questions are categorized into four task-oriented categories based on general research principles~\cite{doi:10.3102/00346543054003327}: study subject \& experimental setup, data characteristics \& collection, technical approach \& details, conclusions \& results.
Details about question setup in Appendix~\ref{subsec:reaserch_Q}

\paragraph{Creation of Answer Choices}
For each question and its paired paper, six answer options are created by domain experts, including one correct answer and five distractors.
The six options consistently include ``All of the above'' and ``None of the above'', either of which may serve as the correct answer when applicable.
Distractors are manually constructed based on: 1) concepts, numerical values, or textual information from the paper that closely resemble the correct answer; or 2) commonly used information or conventions in the respective field.

\section{Experiments}\label{sec:exp}

Our experiments aim to address the following research questions:
(1) \textbf{RQ1} (Section~\ref{sec:Main Result}): Can \agentname effectively solve the \taskname task compared with the baselines?
(2) \textbf{RQ2} (Section~\ref{sec:Answer-Mapping-and-Ablation-Studies}): Do our designed components for \agentname, such as hierarchy preservation, the confidence level mechanism, and the information sufficiency check, function as intended?
(3) \textbf{RQ3} (Section~\ref{subsec:OOD}): Is \agentname robust to variations in evaluation and input conditions, such as different mapping models and non-standard headings?

\subsection{Experiment Settings}\label{subsec: exp setup}

\paragraph{Dataset and Evaluation Metrics}
We use our proposed \benchname for evaluation.
Following Section~\ref{subsec:bench_evaluation}, we use GPT-4.1 to map each short answer generated by the method to one of the predefined answer choices.
Evaluation results using alternative mapping models are presented in Section~\ref{para:sufficiency_check_absent}.
Our primary evaluation metric is accuracy after the answer mapping.

\paragraph{Baseline}

We evaluate a broad range of retrieval-augmented generation (RAG)–based approaches widely adopted for CQA and scientific reasoning, including (1) \textbf{vanilla RAG} using embedding models \textit{all-MiniLM-L6-v2}~\cite{wang2020minilmdeepselfattentiondistillation}, \textit{E5-mistral-7b-instruct}~\cite{mistral}, and \textit{GritLM-7B}~\cite{GritLM}, respectively, with cosine-similarity retrieval over 500-token chunks (50-token overlap); (2) \textbf{contextual RAG} variants that enhance retrieval via dynamic chunk selection and adaptive context expansion~\cite{anthropic2024contextual}; and (3) \textbf{advanced RAG extensions} such as DRAGIN~\cite{suDRAGINDynamicRetrieval2024}, R²AG~\cite{ye_r2ag_2024}, and LongRAG~\cite{jiang2024longragenhancingretrieval}, which introduce multi-hop reasoning, re-ranking, and long-context retrieval.

We also consider representative literature-focused agents, including LUMOS~\cite{yinAgentLumosUnified2024}, PaperQA2~\cite{PaperQA2}, Agentic-Hybrid-RAG~\cite{AgenticHybridRag}, and SciMaster~\cite{chai2025scimaster}, the last of which reports a score of 32.1 on Humanity’s Last Exam, marking the state of the art of scientific agents.

All baselines follow the same \benchname evaluation protocol described above.

Default hyperparameters are used unless otherwise specified; detailed configurations are provided in Appendix~\ref{sec:Baseline Evaluation}.

\begin{table*}[t!]

  \centering

    \small  
\setlength{\tabcolsep}{5.5pt}
\resizebox{\textwidth}{!}{

\begin{tabular}{@{}p{0.6cm}lccccccc@{}}
\toprule
&\textbf{Method} & \textbf{GPT-4o} & \textbf{GPT-4.1} & \textbf{DS-R1} & \textbf{o3}& \textbf{o4-mini} & \textbf{Gemini-2.5 Pro} & \textbf{Llama-3.1-70B} \\
\hline
RAG & Vanilla RAG all-MiniLM-L6-v2 & 60.3 & 61.2 & 64.3 & 60.4 & 61.5 & 61.8 & 59.2\\
& Vanilla RAG E5-mistral-7b-instruct & 59.4 & 64.2 & 63.8 & 60.3 & 61.4 & 59.9 & 60.5\\
& Vanilla RAG GritLM-7B & 60.4 & 63.2 & 63.2 & 59.7 & 58.4 & 58.4 & 61.4\\
& Context. RAG E5-mistral-7b-instruct & 60.7 & 63.8 & 62.8 & 59.1 & 58.3 & 58.9 & 58.9\\
& Context. RAG GritLM-7B & 60.8 & 62.8 & 61.6 & 58.4 & 60.7 & 61.6 & 59.2\\
& DRAGIN & 42.5 & 44.6 & 46.9 & 44.0 & 46.9 & 45.9 & 45.4\\
& R$^{2}$AG & 59.4 & 59.5 & 61.5 & 56.6 & 55.3 & 55.6 & 56.1 \\
& LongRAG & 62.1 & 64.7 & 65.5 & 57.0  & 58.3 & 57.1 &57.4 \\
\midrule
Agent & LUMOS & 50.2 & 52.1 & 55.4 & 55.2 & 56.4 & 54.9 &54.4  \\
& PaperQA2 & 47.7 & 48.9 & 54.0 & 51.8 & 49.2 & 51.2 & 53.8 \\
& Agentic-Hybrid-RAG & 59.8 & 60.2 & 62.3 & 57.5 & 57.8 & 57.2 &56.6 \\
& SciMaster & 59.0 & 57.6 & 63.3 & 57.2 & 58.1 & 57.2 & 57.0 \\
& \textbf{\agentname (Ours)} & \textbf{70.0} & \textbf{75.8} & \textbf{74.4} & \textbf{73.4} & \textbf{73.8} & \textbf{75.9} & \textbf{68.8}\\
\bottomrule
\end{tabular}
}
\caption{Cross-domain accuracy (in \%, defined by the macro average over the five domains) on \benchname. Our \agentname uniformly outperforms the RAG-based retrieval and agent-based baselines across the five domains for seven model choices. See Appendix \ref{subsec: breakdown} for the complete breakdown results.}
\label{tab:model-performan}
\end{table*}

\begin{figure*}[t!]

  \centering
  \includegraphics[width=1\linewidth]{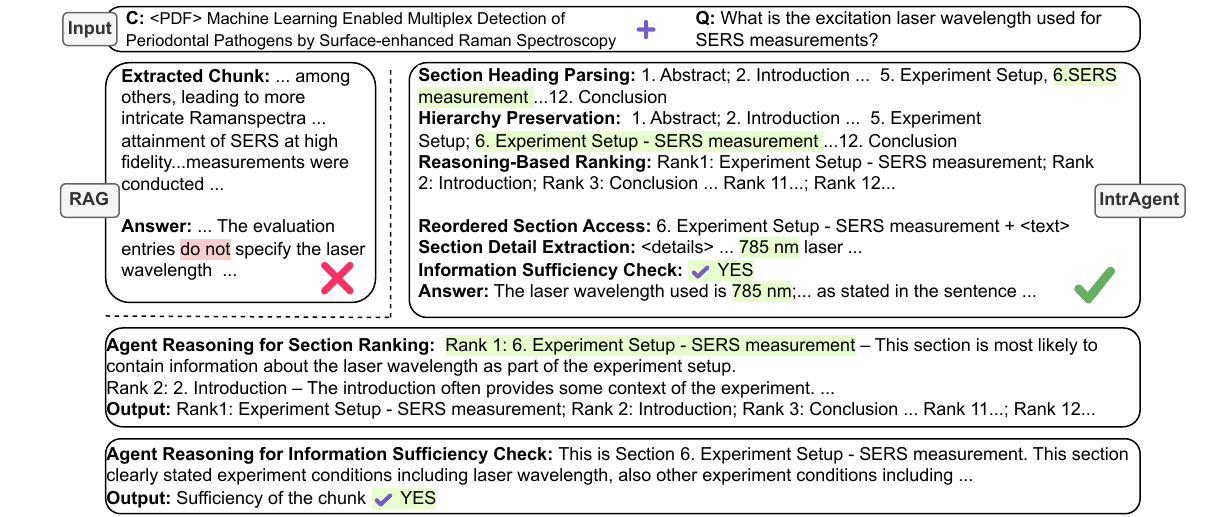}
  \caption{An example of \agentname executing a question-paper pair from \benchname.
  For an input question $Q$ regarding paper $C$, vanilla RAG fails to extract the correct chunk, resulting in an incorrect answer.
  In contrast, \agentname ranks the sections through reasoning and retrieves the correct details that pass the sufficiency check in the first iteration, leading to a correct answer.
  Details for section ranking and sufficiency check are also presented.
  }
  \label{fig2}
\end{figure*}

\subsection{Main Result}
\label{sec:Main Result}

Table \ref{tab:model-performan} show that \agentname sets a new state of the art across all five domains of \benchname for seven backbone LLMs.
Averaged over physics, public health, earth science, engineering, and material science, \agentname achieves 70.0\% with GPT-4o, 75.8\% with GPT-4.1, 74.4\% with DeepSeek-R1, 73.4\% with o3, 73.8\% with o4-mini, 75.9\% with Gemini-2.5 Pro, and 68.8\% with Llama-3.1-70B.
Compared with the strongest baseline under \textit{each} model, \agentname surpasses them by 7.9\%, 11.6\%, 8.9\%, 18.2\%, 17.4\%, 21.0\%, and 7.4\%, respectively.
These results suggest that performance gains stem not from increased context length, but from targeted evidence selection: unlike chunk-based methods that introduce irrelevant context, IntrAgent isolates task-relevant sections, enabling more precise reasoning.

Such a performance gain stems from our novel agent design: 1) hierarchy‑aware section ranking based on reasoning, and 2) sufficiency check that halts reading once evidence is complete.

In contrast, RAG sequentially feeds multiple unstructured text chunks into the LLM, which often introduces irrelevant or noisy information.
The agent baselines, originally designed for scientific QA through online search, degenerate into static information retrieval pipelines similar to RAG when they are directly provided with the literature for \taskname.

A representative example demonstrating the advantage of \agentname over the RAG baseline is shown in Figure~\ref{fig2}, along with the reasoning trajectory for \agentnamenospace’s section ranking and sufficiency check. 
In this example, the agent prioritizes the Experiment Setup – SERS measurement section based on the reasoning that laser wavelength information is typically part of the experimental setup. During the information sufficiency check, this section is verified to explicitly state the excitation wavelength (785 nm), confirming that the relevant detail is correctly retrieved from the appropriate section.
These results address \textbf{RQ1}, confirming that \agentname more effectively solves the \taskname task compared to the baselines.

\subsection{Ablation Studies}
\label{sec:Answer-Mapping-and-Ablation-Studies}

\paragraph{Study on Hierarchy Preservation}
We evaluate \agentname without the hierarchy preservation step by directly ranking the raw section headings $\mathcal{H}_0$ extracted from the Markdown-parsed text.
As shown in Table~\ref{tab:model-domain-perform_no_tree}, removing the hierarchy preservation step leads to a clear drop in cross-domain accuracy for all three model choices, highlighting the necessity of this key step for effectively incorporating structural context into section ranking.

\begin{table}[t]
\small
  \centering
  \small
  \setlength{\tabcolsep}{4.5pt}
  \begin{tabular}{lccc}
    \toprule
    Model & GPT-4o & GPT-4.1 & DeepSeek-R1 \\
    \midrule
    w/o HP & 60.7 & 70.3 & 64.2 \\
    w/ HP  & \textbf{65.6} & \textbf{72.5} & \textbf{67.6} \\
    \bottomrule
  \end{tabular}
    \caption{Accuracy (\%) of \agentname{} w/ and w/o the hierarchy preservation (HP) step for three model choices.}
    \label{tab:model-domain-perform_no_tree}
\end{table}

\paragraph{Study on Confidence Level} 
\agentname supports three confidence levels via prompt variation during iterative reading: 1) Conservative, which halts only when all answer components are explicitly observed; 2) Balanced (the default), which stops when evidence appears sufficient; and 3) Aggressive, which allows early termination based on partial evidence.
The full prompts are provided in Appendix~\ref{prompt:Confidence-Level}.
This confidence level parameter enables users of \agentname to control the operational overhead (more details on overhead are shown in Appendix~\ref{subsec: run_time}).
As shown in Table~\ref{tab:confidence-ablation}, more aggressive reading reduces the number of iterative reading steps but at the cost of lower information retrieval accuracy.
Interestingly, the conservative mode, despite reading more sections, performs the worst, consistent with RAG observations that performance degrades with very long context windows~\cite{jiang2024longragenhancingretrieval,yepes2024financialreportchunkingeffective,wang2024searchingbestpracticesretrievalaugmented}.

\begin{table}[t]
\small
  \centering
  \small
  \setlength{\tabcolsep}{2.4pt}
  \begin{tabular}{lccc}
    \toprule
    Metric & Conservative & Balanced & Aggressive \\
    \midrule
    Accuracy (\%)        & 58.9 & 68.3 & 62.7 \\
    Avg. Iterations      & 9.9  & 5.1  & 3.9 \\
    Med. Iterations      & 11   & 2    & 1   \\
    Std Dev. Iterations  & 7.9  & 5.5  & 5.3 \\
    Med. Token Count & 7853 & 6376 & 2233\\
    \bottomrule
  \end{tabular}
    \caption{Impact of confidence level on the MCQ accuracy, the number of reading iterations, and token count.}
    \label{tab:confidence-ablation}
\end{table}

\paragraph{Study on Information Sufficiency Check}\label{para:sufficiency_check_absent}
We evaluate \agentname without the information sufficiency check by accessing only the top-1 section during iterative reading.
On the physics dataset, using GPT-4o as the backbone LLM and GPT-4.1 as the mapping model, this modification results in a substantial accuracy drop from 75.4\% to 32.2\%, showing the necessity of this critical component.
Moreover, analysis of failure cases reveals two key issues:
1) Incomplete retrieval: 
The necessary information could be distributed across multiple sections or require cross-referencing.
Although the most relevant section is correctly identified in most cases, it alone may not suffice to answer the query -- leading the agent to respond with ``None of the above.''
2) Hallucination: The agent produces confident but unsupported answers.
An example failure case for hallucination is provided in Appendix~\ref{subsec: Failure Case}.

These results confirm that all the designed components function as intended, addressing \textbf{RQ2}.

\begin{table}[t]
  \centering
  \small
  \setlength{\tabcolsep}{4pt}
  \begin{tabular}{lcc|c}
    \toprule
    \textbf{Mapping} & \textbf{\agentname} & \textbf{RAG} & \textbf{Avg.} \\
    \midrule
    GPT-4o      & 59/65 & 60/65 & 59.5/65 \\
    GPT-4.1     & 65/65 & 61/65 & \textbf{63/65} \\
    DeepSeek-R1 & 61/65 & 61/65 & 61/65 \\
    \bottomrule
  \end{tabular}
    \caption{Correctness agreement between human annotations and LLM-based mapping on the physics dataset with 65 instances.}
    \label{tab:human-compare}
\end{table}

\subsection{Robustness of \agentname to Mapping Model and Input Variability}\label{subsec:OOD}

\paragraph{Impact of Mapping Model}
Since the evaluation protocol of \benchname involves mapping short-form answers to multiple-choice (MCQ) selections, it is important to assess the reliability of the mapping model.

Using identical prompts on the physics subset, we compare the mappings produced by GPT-4o, GPT-4.1, and DeepSeek-R1 against ground-truth annotations provided by domain experts.
Here, we consider short-form answers generated by \agentname (with GPT-4o as the backbone) and by vanilla RAG.
We also define a \textit{correctness agreement} metric as the ratio of MCQ choices that match the human label in correctness.

As shown in Table~\ref{tab:human-compare}, GPT-4o and GPT-4.1 yield nearly identical accuracy, with GPT-4.1 achieving slightly better alignment (63 out of 65).
Based on these results, we adopt GPT-4.1 as the default mapping model for all main experiments.
We hypothesize that LLMs with stronger scientific reasoning and domain expertise could further improve mapping quality—an investigation we leave to future work.

We also conduct a parallel study on the same physics dataset, where short-form answers are generated by \agentname using three different backbone models, as well as by the vanilla RAG baseline.
Each answer is then mapped to an MCQ option using the same prompt across three different mapping models.
As shown in Table~\ref{tab:mapping-compare}, although accuracy varies across backbone-mapping model combinations, \agentname consistently outperforms the vanilla RAG baseline, which is the strongest among all baselines in our experiments.

\begin{table}[t]
  \centering
  \small
  \setlength{\tabcolsep}{2pt}
  \begin{tabular}{lcccc}
    \toprule
    & \multicolumn{3}{c}{\textbf{\agentname}} & \\
    \cmidrule(lr){2-4}
    \textbf{Mapping} & \textbf{GPT-4o} & \textbf{GPT-4.1} & \textbf{DeepSeek-R1} & \textbf{RAG} \\
    \midrule
    GPT-4o      & 73.8 & 76.9 & 73.8 & 63.1 \\
    GPT-4.1     & 75.4 & 78.5 & 72.3 & 60.0 \\
    DeepSeek-R1 & 67.7 & 70.8 & 76.9 & 61.5 \\
    \bottomrule
  \end{tabular}
    \caption{Accuracy (\%) of \agentname{} for different mapping models compared with the Vanilla RAG baseline on the physics dataset.}
      \label{tab:mapping-compare}
\end{table}
\paragraph{Robustness of \agentname Under Subpar Section Headings}
Here, we investigate an interesting edge case where a research paper fails to annotate its sections with standard headings. 
We select a representative paper from the physics dataset~\cite{rathnayake_machine_2024} and construct three alternative heading styles for the paper: a beginner-style rewrite, a highly noisy version, and a Shakespearean rendition. The full list of rewrites is shown in Appendix~\ref{appx:bad_title}.
With GPT-4o as the backbone, \agentname achieves 89.2\% accuracy on the original headings.
For all three altered versions, it maintains strong performance, achieving 84.6\% accuracy with only a modest drop.
Appendix~\ref{appx:bad_title_exp} provides detailed reasoning traces illustrating how \agentname effectively handles noisy headings during section ranking.

These results address \textbf{RQ3}, demonstrating the robustness of \agentname to variations in both evaluation settings and input conditions.

\section{Conclusion}

We introduce \taskname, a novel task that targets the critical yet underexplored practice of retrieving information from scientific literature.
We introduce \agentname, the first LLM-based agent specifically designed for \taskname.
\agentname mimics human reading behavior by first identifying the most relevant sections and then iteratively extracting key information.
To enable systematic evaluation of \taskname approaches, including \agentname, we present \benchname -- a benchmark spanning five high-impact scientific domains.
Experimental results show the superior performance of \agentname over baselines on this benchmark.

\section*{Limitations}
While \agentname and \benchname focus on advancing text-based scientific information retrieval and understanding, this work does not yet incorporate non-textual modalities such as plots, figures, and tables. These visual elements often encapsulate concentrated insights, including experimental trends, quantitative comparisons, and structural relationships, which can be essential for comprehensive scientific reasoning.

In addition, the literature considered in this paper encompasses the major article types, but not all. For example, review papers are not included in our evaluation.
In the future work, we will expand our benchmark to include more types of papers and queries.

\bibliography{references}

\clearpage
\appendix
\onecolumn

\twocolumn                
\section{Details about \benchname}
\label{sec: Details about the dataset}

The benchmark was created through a systematic process: first selecting representative science literature, then manually crafting questions that span different scientific four task-oriented categories: study subject \& experimental setup, data characteristics \& collection, technical approach \& details, conclusions \& results. Special attention was given to ensuring professional quality and coverage across multiple subfields.

\subsection{Research Fields in \benchname}\label{subsec:Research Fields in IntraBench}

\paragraph{Surface-enhanced Raman Spectroscopy (SERS)} \textit{Surface-enhanced raman spectroscopy}(SERS) represents a significant advancement in applied physics, particularly within the domain of optical physics. At its core, SERS is an extension of Raman spectroscopy, a technique that analyzes the inelastic scattering of light to provide insights into molecular vibrations. The enhancement in SERS arises when molecules are adsorbed onto nanostructured metallic surfaces, such as gold or silver, leading to a substantial increase in the Raman signal. This amplification is primarily attributed to localized surface plasmon resonances—coherent oscillations of conduction electrons at the metal surface excited by incident light—which generate intense electromagnetic fields near the surface, thereby boosting the Raman scattering efficiency of nearby molecules. The remarkable sensitivity of SERS enables the detection of analytes at extremely low concentrations, down to the single-molecule level. This capability is particularly valuable in various engineering applications where identifying trace amounts of substances is crucial. For instance, in public safety, SERS-based sensors have been developed for the rapid and accurate detection of bio-hazardous materials, including chemical agents, viruses, and bacteria\cite{thrift_quantification_2019,rathnayake_machine_2024,yang_rapid_2024,yang_rapid_2023,erzina2022quantitative}. Such applications benefit from SERS's ability to provide specific molecular fingerprints, facilitating the identification of dangerous substances even in complex environments.

\paragraph{Remote Sensing} Land cover and land use classification is a central research direction in remote sensing science, aiming to automatically identify and spatially represent surface types and human activity patterns based on multi-source remote sensing data\cite{gong_annual_2020,liu_annual_2020,gong_finer_2013,li_mapping_2021,gong_mapping_2020}. This field widely utilizes medium- and high-resolution satellite imagery from optical sensors (e.g., Landsat, Sentinel-2) and synthetic aperture radar (e.g., Sentinel-1), employing techniques such as supervised classification, object-based analysis, time-series processing, and deep learning for high-accuracy mapping. In recent years, deep convolutional neural networks such as U-Net, DeepLab, and HRNet have been extensively applied to semantic segmentation of high-resolution imagery, significantly improving the extraction accuracy of fine-scale urban features such as buildings, roads, and impervious surfaces. In addition, the integration of optical and radar data with auxiliary variables—such as nighttime lights and population density—has enhanced model performance in complex urban environments and facilitated dynamic monitoring of land use changes. These studies are of great value for supporting global change research, environmental assessment, and sustainable urban planning.

\paragraph{Infectious‐disease Modeling} Infectious‐disease modeling and forecasting play a central role in public health by enabling researchers and policymakers to anticipate the spread and burden of infectious diseases. Among the tools available, mathematical modeling—particularly compartmental models such as SIR and SEIR—has proven indispensable for simulating disease transmission dynamics, assessing intervention strategies, and guiding resource allocation\cite{khan_mathematical_2022,biswas2020covid,anggriani2022mathematical,iboi_mathematical_2020,ndairou_mathematical_2020}. The COVID-19 pandemic has exemplified the value of these models, as they have been critical for estimating infection trajectories, evaluating the timing and effectiveness of control measures, and informing real-time policy responses during a rapidly evolving global crisis. Accurate modeling has thus been essential for understanding and managing the impact of COVID-19 across different settings and timeframes.

\paragraph{Human Factor} Human factors is a cornerstone of industrial engineering, focusing on the design and evaluation of systems that balance human well-being and overall system performance. Human factors has addressed challenges in physical workload, cognitive workload, and human–machine interaction across domains such as manufacturing, transportation, and healthcare. A critical area is the classification and prediction of human fatigue and risk. This is because prolonged exposure to repetitive motions, awkward postures, or high-intensity workloads can impair human performance, increase error rates, and elevate injury risk. Recent studies have leveraged wearable sensors, biomechanical modeling, and machine learning methods to detect early indicators of fatigue and quantify risks, enabling proactive intervention\cite{lamooki2022data,escobar2022worker,ansari2022automatic,ding2025assessing,su2024enhancing}. Such approaches underscore the evolving role of human factors research in advancing resilient, human-centered industrial systems.

\paragraph{Additive Manufacturing} Additive manufacturing (AM), rooted in materials science and manufacturing engineering, refers to a family of processes that build components layer by layer directly from digital models. Over the past decades, AM has evolved from rapid prototyping into a transformative production technology, enabling complex geometries, lightweight structures, and customized products across aerospace, biomedical, and energy sectors. While the promise of AM lies in its design freedom and material efficiency, ensuring consistent quality remains a central challenge. Quality control in AM encompasses detecting defects such as porosity and dimensional inaccuracy, which can critically impact the structural integrity and performance of printed parts. Researchers increasingly employ in-situ monitoring, computer vision, and machine learning-based defect detection to address variability AM processes\cite{zhang2024autonomous,li2021geometrical,seifi2019layer,segura2021online,gaikwad2020toward}. Such quality-control efforts are pivotal to advancing AM toward reliable, industrial-scale deployment.

\subsection{Research Literature in \benchname}

Table~\ref{tab:paper_table} contains all 25 papers, five per field, used in \benchname. 

\begin{table*}[htbp]
\centering

\small
\begin{tabular}{p{11cm}r}
\toprule
\textbf{Title} & \textbf{Ref.} \\
\midrule

\multicolumn{2}{@{}l}{\textbf{Public Health} - Infectious-disease Modeling}  \\[4pt]
Mathematical modeling and analysis of COVID-19: a study of new variant Omicron & \cite{khan_mathematical_2022}\\
COVID-19 pandemic in India: a mathematical model study & \cite{biswas2020covid} \\
A mathematical COVID-19 model considering asymptomatic and symptomatic classes with waning immunity & \cite{anggriani2022mathematical} \\
Mathematical modeling and analysis of COVID-19 pandemic in Nigeria & \cite{iboi_mathematical_2020} \\
Mathematical modeling of COVID-19 transmission dynamics with a case study of Wuhan & \cite{ndairou_mathematical_2020} \\

\midrule
\addlinespace
\multicolumn{2}{@{}l}{\textbf{Physics} - Surface Enhanced Raman Spectroscopy} \\[4pt]
Quantification of analyte concentration in the single molecule regime using convolutional neural networks & \cite{thrift_quantification_2019} \\
Machine learning enabled multiplex detection of periodontal pathogens by surface-enhanced Raman spectroscopy & \cite{rathnayake_machine_2024} \\
Rapid detection of SARS-CoV-2 variants using an angiotensin-converting enzyme 2-based surface-enhanced Raman spectroscopy sensor enhanced by CoVari deep learning algorithms & \cite{yang_rapid_2024} \\
Rapid detection of SARS-CoV-2 RNA in human nasopharyngeal specimens using surface-enhanced Raman spectroscopy and deep learning algorithms & \cite{yang_rapid_2023} \\
Quantitative detection of $\alpha$1-acid glycoprotein (AGP) level in blood plasma using SERS and CNN transfer learning approach & \cite{erzina2022quantitative} \\

\midrule
\addlinespace
\multicolumn{2}{@{}l}{\textbf{Earth Science} - Remote Sensing} \\[4pt]
Annual maps of global artificial impervious area (GAIA) between 1985 and 2018 & \cite{gong_annual_2020} \\
Annual dynamics of global land cover and its long-term changes from 1982 to 2015 & \cite{liu_annual_2020}\\
Finer resolution observation and monitoring of global land cover: first mapping results with Landsat TM and ETM+ data & \cite{gong_finer_2013} \\
Mapping essential urban land use categories in Beijing with a fast area of interest (AOI)-based method & \cite{li_mapping_2021} \\
Mapping essential urban land use categories in China (EULUC-China): preliminary results for 2018 & \cite{gong_mapping_2020} \\

\midrule
\addlinespace
\multicolumn{2}{@{}l}{\textbf{Engineering} - Human Factor} \\[4pt]
A data analytic end-to-end framework for the automated quantification of ergonomic risk factors across multiple tasks using a single wearable sensor & \cite{lamooki2022data} \\
Assessing human situation awareness reliability considering fatigue and mood using EEG data: a Bayesian neural network-Bayesian network approach & \cite{ding2025assessing}\\
Automatic driver cognitive fatigue detection based on upper body posture variations  & \cite{ansari2022automatic} \\
Enhancing data privacy in human factors studies with federated learning & \cite{su2024enhancing} \\
Worker’s physical fatigue classification using neural networks  & \cite{escobar2022worker} \\

\midrule
\addlinespace
\multicolumn{2}{@{}l}{\textbf{Material Science} - Additive Manufacturing} \\[4pt]
Autonomous optimization of process parameters and in-situ anomaly detection in aerosol jet printing by an integrated machine learning approach & \cite{zhang2024autonomous} \\
Geometrical defect detection for additive manufacturing with machine learning models & \cite{li2021geometrical}\\
Layer-wise modeling and anomaly detection for laser-based additive manufacturing & \cite{seifi2019layer} \\
Online droplet anomaly detection from streaming videos in inkjet printing & \cite{segura2021online} \\
Toward the digital twin of additive manufacturing: integrating thermal simulations, sensing, and analytics to detect process faults & \cite{gaikwad2020toward} \\

\bottomrule
\end{tabular}
\caption{Literature across five scientific domains used in our benchmark.}
\label{tab:paper_table}
\end{table*}

\subsection{Research Questions in \benchname}\label{subsec:reaserch_Q}

In total, \benchname comprises 63 research questions, systematically categorized into four task-oriented categories: (1) Study subject \& experimental setup, (2) Data characteristics \& collection, (3) Technical approach \& details, and (4) Conclusions \& results. These questions are designed to elicit critical information from scientific papers across five domains. Specifically, the \textit{SERS in chemistry physics} dataset contributes 13 questions, the \textit{infectious‐disease modeling} in public health dataset contributes 11 questions, the \textit{remote sensing} in earth science dataset contributes 12 questions, the \textit{human performance sensing} in engineering dataset contributes 13 questions, and the \textit{additive manufacturing} in material science dataset contributes 14 questions. The complete lists are provided in Table~\ref{tab:phys_q}, Table~\ref{tab:ph_q}, Table~\ref{tab:es_q}, Table~\ref{tab:eng_q}, and Table~\ref{tab:ms_q}, respectively.

Across all domains, the 63 questions remain balanced across the first three task-oriented categories—Study subject \& experimental setup, Data characteristics \& collection, and Technical approach \& details—each covering foundational aspects of experimental design, data acquisition, and analytical methodology. The remaining questions belong to the Conclusions \& results category, emphasizing the outcome evaluation and performance reporting aspects of scientific research.

While the overall structure 63 consistent, each domain reflects distinct emphases. The \textit{infectious‐disease modeling} dataset prioritizes experimental setup due to the importance of compartmental structures and intervention parameters. The \textit{remote sensing} dataset highlights data characteristics such as sensor type, spatial resolution, and temporal coverage. The \textit{SERS in physics} dataset shows a balanced distribution aligned with typical SERS workflows from substrate preparation to data-driven analysis. The \textit{humanfactor in engineering} dataset emphasizes human performance measurement and multimodal sensing setups, reflecting the diversity of sensor placements and physiological variables in human studies. The \textit{Additive manufacturing in material science} dataset focuses on process monitoring and machine learning evaluation within additive manufacturing workflows, emphasizing material–process–defect relationships and reproducible modeling strategies.

To ensure usability and reproducibility, each research question was expert-annotated with detailed explanatory notes.

\begin{table*}[h]
  \centering
    \small 
  \begin{tabular}{p{3cm} p{10cm}}
    \toprule
    \textbf{Task-oriented Category} & \textbf{Research Question $Q$} \\
    \midrule
    \multirow{3}{=}{Study subject \& experimental setup} & What are the main analytes type studied? \\
    & What are the material and structure, or morphology of the SERS substrates used?\\
     & How many analytes are investigated? \\

    \midrule
    \multirow{4}{=}{Data characteristics \& collection}  & What is the excitation laser wavelength used for SERS measurements? \\
    & What is the spectral range collected for the analysis of the analytes? \\
     & How many spectra are collected per analyte under each experimental condition? \\
     & How many experimental replications are conducted to ensure reproducibility? \\
     
    \midrule
    \multirow{4}{=}{Technical approach \& details} & What is the primary machine learning task addressed in this study? \\
     & Which machine learning algorithm is implemented? \\
     & What data splitting strategy is applied, and the parameters?  \\
     & How many epochs are used during model training?  \\
     
    \midrule
    \multirow{2}{=}{Conclusions \& results} & What performance metrics are employed to evaluate the machine learning models?\\
     & What are the reported performance values? \\
    \bottomrule
  \end{tabular}
      \caption{Research Questions in \benchname: \textit{SERS in chemistry physics} (Phys)}
       \label{tab:phys_q}
\end{table*}

\begin{table*}[h]
  \centering

    \small 
  \begin{tabular}{p{3cm} p{10cm}}
    \toprule
    \textbf{Task-oriented Category} & \textbf{Research Question $Q$} \\
    \midrule
    \multirow{4}{=}{Study subject \& experimental setup} & Into how many compartments is the population divided in the model? \\
     & What is the initial susceptible population of the model? \\
     & What is the initial infected population of the model? \\
     & How many interventions are addressed in the paper? \\

    \midrule
    \multirow{2}{=}{Data characteristics \& collection} & What is the source location or country of origin for the data used in this study? \\
    \\

    \midrule
    \multirow{4}{=}{Technical approach \& details} & What is the model used in this paper? \\
     & What is the transmission rate? \\
     & What is the disease-induced mortality rate? \\
     & What values of the basic reproduction number were considered in the model? \\

    \midrule
    \multirow{2}{=}{Conclusions \& results} & What are the novel contributions of the paper? \\
     & What are the limitations of the paper? \\
    \bottomrule
  \end{tabular}
      \caption{Research Questions in \benchname: \textit{infectious‐disease modeling} in Public Health (PH)}
          \label{tab:ph_q}
\end{table*}

\begin{table*}[h]
  \centering
    \small 
  \begin{tabular}{p{3cm} p{10cm}}
    \toprule
    \textbf{Task-oriented Category} & \textbf{Research Question $Q$} \\
    \midrule
    \multirow{3}{=}{Study subject \& experimental setup} & What is the number of land-cover / land-use classes classified in this study? \\
     & What is the spatial extent of the study area? \\
     & What is the geographic type of the study area? \\

    \midrule
    \multirow{5}{=}{Data characteristics \& collection} & What is the temporal scope of the data used? \\
     & What type of remote sensing data is used? \\
     & Which specific satellite data is used? \\
     & What is the spatial resolution of the primary imagery used? \\
     & Are auxiliary features used beyond raw spectral bands? \\

    \midrule
    \multirow{2}{=}{Technical approach \& details} & What type of model is implemented in this study? \\
     & What performance metrics are reported? \\

    \midrule
    \multirow{2}{=}{Conclusions \& results} 
     & Is any comparative analysis included? \\
& What is the reported overall accuracy (OA)? \\
    \bottomrule
  \end{tabular}
    \caption{Research Questions in \benchname: \textit{remote sensing} in Earth Science (ES)}
    \label{tab:es_q}
\end{table*}

\begin{table*}[h]
  \centering
    \small 
  \begin{tabular}{p{3cm} p{10cm}}
    \toprule
    \textbf{Task-oriented Category} & \textbf{Research Question $Q$} \\
    \midrule
    \multirow{3}{=}{Study subject \& experimental setup} 
      & What are the subjects’ occupational roles? \\
      & What specific task or activity are the subjects performing? \\
      & What is the study context or environment in which participants perform the tasks? If the data are referenced from prior work, please indicate the source. \\

    \midrule
    \multirow{4}{=}{Data characteristics \& collection}  
      & What are the primary sensing modalities or measurement instruments employed in the study to capture human performance and physiological responses? \\
      & What is the anatomical or body placement of the sensors used in the study? \\
      & What is the sampling rate (Hz)? \\
      & What is the total number of participants involved in the study? \\

    \midrule
    \multirow{4}{=}{Technical approach \& details} 
      & How are the participants’ physical, cognitive, or perceptual states assessed or reflected in the study? \\
      & What is the primary modeling objective in this study? \\
      & What is the data partitioning strategy used during model training, and what are the parameters? \\
      & What is the number of epochs used during model training (i.e., how many complete passes through the entire training dataset)? \\

    \midrule
    \multirow{2}{=}{Conclusions \& results} 
      & Which performance metrics are used to assess the effectiveness of the machine learning models? \\
      & What are the reported values for the performance metrics used to evaluate the machine learning models? \\
    \bottomrule
  \end{tabular}
      \caption{Research Questions in \benchname: \textit{Human performance sensing in engineering} (Engr)}
      \label{tab:eng_q}
\end{table*}

\begin{table*}[h]
  \centering
    \small 
  \begin{tabular}{p{3cm} p{10cm}}
    \toprule
    \textbf{Task-oriented Category} & \textbf{Research Question $Q$} \\
    \midrule
    \multirow{4}{=}{Study subject \& experimental setup} 
      & What type of additive manufacturing process is studied? \\
      & What type of material is used for printing? \\
      & What kind of shape or product is printed? \\
      & What is the primary defect being studied? \\

    \midrule
    \multirow{3}{=}{Data characteristics \& collection}  
      & What sensors are used to measure the process? \\
      & What is the sampling rate (Hz)? \\
      & If relevant, what is the spatial resolution (µm)? \\

    \midrule
    \multirow{5}{=}{Technical approach \& details} 
      & What is the machine learning objective in this study? \\
      & What machine learning algorithm is used? \\
      & How are the data split during machine learning? \\
      & How many replications are conducted during machine learning, if any? \\
      & How many epochs are used during machine learning, if any? \\

    \midrule
    \multirow{2}{=}{Conclusions \& results} 
      & What metrics are used to evaluate the machine learning models? \\
      & What are the values for these metrics? \\
    \bottomrule
  \end{tabular}
    \caption{Research Questions in \benchname: \textit{Additive manufacturing in material science} (MS)}
    \label{tab:ms_q}
\end{table*}

\subsection{Main Result Report over Task-oriented Categories}

To further analyze performance differences among backbone large language models (LLMs), we evaluate \agentname\ across four \textit{task-oriented categories}: (1) \textit{Study subject \& experimental setup (S\&E)}, (2) \textit{Data characteristics \& collection (D\&C)}, (3) \textit{Technical approach \& details (T\&D)}, and (4) \textit{Conclusions \& results (C\&R)}.

\begin{figure}[t!]
\centering
\includegraphics[width=0.85\linewidth]{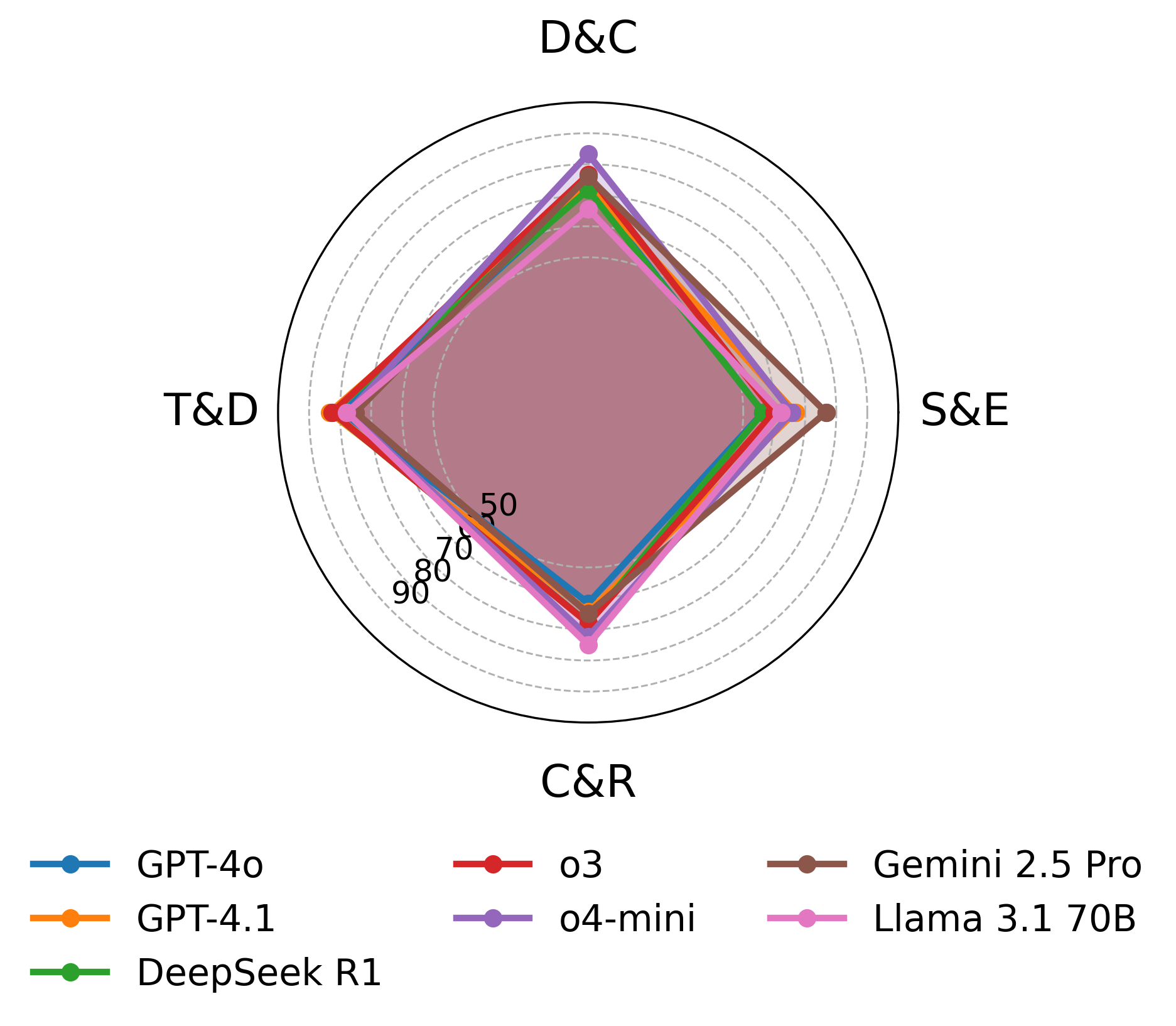}
\caption{Radar plot comparing the performance of seven backbone LLMs—GPT-4o, GPT-4.1, DeepSeek-R1, o3, o4-mini, Gemini 2.5 Pro, and Llama-3.1-70B—across four research-question categories: Study subject \& experimental setup (S\&E), Data characteristics \& collection (D\&C), Technical approach \& details (T\&D), and Conclusions \& results (C\&R). GPT-4.1 and o4-mini show the strongest overall balance, while Gemini 2.5 Pro and Llama-3.1-70B demonstrate stable domain-specific strengths. The plot highlights that despite moderate variation across backbones, \agentname\ remains consistently robust across all task categories.}
\label{radar}
\end{figure}

These evaluations reveal how each backbone contributes to different aspects of literature reasoning. Figure~\ref{radar} visualizes the results across seven LLMs. Two overall trends emerge.  
First, \agentname\ maintains balanced performance across all categories regardless of the backbone, confirming its robustness and model-agnostic design.  
Second, although individual models show slight strengths in specific areas, the overall variation remains moderate, indicating that the contextual reading and iterative synthesis strategy of \agentname\ mitigates backbone dependency.

In detail, GPT-4.1 achieves the highest overall balance (S\&E = 66.8, D\&C = 72.7, T\&D = 83.4, C\&R = 64.3), especially excelling in technical reasoning.  
o4-mini performs comparably (65.7, 83.3, 77.5, 72.3), outperforming GPT-4.1 in D\&C and C\&R.  
Gemini 2.5 Pro maintains uniformly high results (76.7, 76.0, 75.0, 64.8), reflecting strong generalization.  
DeepSeek-R1 (56.4, 71.8, 82.6, 67.9) and o3 (60.0, 76.7, 82.5, 68.0) show stable interpretative abilities, while Llama-3.1-70B (62.2, 65.6, 77.9, 75.0) exhibits strength in conclusion-oriented reasoning.  

Overall, GPT-4.1 and o4-mini demonstrate the most consistent and well-rounded performance, while Gemini 2.5 Pro and Llama-3.1-70B show domain-specific advantages.  
These results suggest that \agentname’s architecture effectively harmonizes reasoning depth and retrieval accuracy across heterogeneous LLMs, ensuring adaptability without overfitting to any specific model.

\section{Additional Experiments and Reports}\label{sec: more exp}

\subsection{Main Experiment - Domain-level Accuracy Breakdown}\label{subsec: breakdown}

Table~\ref{tab:model-performan} in the main text reports averaged accuracies for every
\textit{baseline–LLM} combinations.  
For each pair we compute the simple arithmetic mean over the five domain datasets shown below, so every domain carries equal weight: \textit{SERS} in physics, \textit{infectious-disease modeling} in public health, \textit{remote sensing} in earth science, \textit{human performance sensing} in engineering, \textit{additive manufacturing} in material science.

While the main table focuses on these aggregated averages, the following tables (Table~\ref {tab:model-performance-phys},~\ref{tab:model-performance-ph},~\ref{tab:model-performance-es},~\ref{tab:model-performance-hf},~\ref{tab:model-performance-mfg}) present the detailed per-domain accuracies for all \textit{baseline–LLM} combinations.  
Each table corresponds to one of the five domains listed above.  
Boldface highlights the best score within each domain, allowing readers to examine where specific retrieval strategies or LLM backbones perform most effectively.

\begin{table*}[t!]
  \centering
    \small
\setlength{\tabcolsep}{3.5pt}
\resizebox{\textwidth}{!}{
\begin{tabular}{@{}p{0.6cm}lccccccc@{}}
\hline
&\textbf{Baseline} & \textbf{GPT-4o} & \textbf{GPT-4.1} & \textbf{DS-R1} & \textbf{o3}& \textbf{o4-mini} & \textbf{Gemini-2.5 Pro} & \textbf{Llama-3.1-70B} \\
\hline
RAG & Vanilla RAG all-MiniLM-L6-v2 & 60.0 & 66.2 & 64.6 & 58.5 & 60.0 & 60.0 & 67.7 \\
 & Vanilla RAG E5-mistral-7b-instruct & 61.5 & 70.7 & 63.8 & 63.1 & 64.6 & 64.6 & 66.2 \\
 & Vanilla RAG GritLM-7B & 60.0 & 67.6 & 58.5 & 58.5 & 60.0 & 58.5 & 64.6 \\
 & Contextual RAG E5-mistral-7b-instruct & 58.5 & 66.2 & 58.5 & 58.5 & 60.0 & 53.9 & 58.5 \\
 & Contextual RAG GritLM-7B & 53.9 & 67.6 & 64.6 & 58.5 & 60.0 & 60.0 & 58.5 \\
 & DRAGIN & 38.5 & 43.0 & 47.7 & 43.1 & 46.2  & 44.6  & 46.2  \\
 & R$^{2}$AG & 58.5 & 56.9 & 58.5 & 56.9  &56.9 & 56.9  & 55.4  \\
 & LongRAG & 64.6 & 70.7 & 64.6 & 56.9  & 58.5  & 55.4  & 56.9 \\
\midrule
Agent & LUMOS & 52.3 & 50.8 & 50.8 & 53.8  & 56.9  &56.9 & 56.9  \\
 & PaperQA2 & 49.2 & 50.8 & 52.3 & 49.2  & 52.3  & 50.8  & 52.3  \\
 & Agentic-Hybrid-RAG & 60.0 & 63.1 & 61.5 & 55.4 & 55.4 & 55.4  & 53.8  \\
 & SciMaster & 58.5 & 58.5 & 63.1 & 53.8  & 58.5  & 55.4  & 55.4  \\
 & \textbf{\agentname (Ours)} & 75.4 & 78.5 & 78.5 & 76.9 & 81.5 & 78.5 & 72.3 \\
\hline
\end{tabular}

}
\caption{Accuracies (\%) on \benchname, Physics
}
\label{tab:model-performance-phys}
\end{table*}

\begin{table*}[t!]
  \centering

    \small
\setlength{\tabcolsep}{3.5pt}
\resizebox{\textwidth}{!}{
\begin{tabular}{@{}p{0.6cm}lccccccc@{}}
\hline
&\textbf{Baseline} & \textbf{GPT-4o} & \textbf{GPT-4.1} & \textbf{DS-R1} & \textbf{o3}& \textbf{o4-mini} & \textbf{Gemini-2.5 Pro} & \textbf{Llama-3.1-70B} \\
\hline
RAG & Vanilla RAG all-MiniLM-L6-v2 & 52.7 & 60.0 & 65.6 & 58.2 & 61.8 & 58.2 & 54.5 \\
 & Vanilla RAG E5-mistral-7b-instruct & 52.7 & 61.7 & 65.6 & 54.6 & 56.4 & 54.6 & 58.2 \\
 & Vanilla RAG GritLM-7B & 52.7 & 61.7 & 58.2 & 56.4 & 52.7 & 54.5 & 58.2 \\
 & Contextual RAG E5-mistral-7b-instruct & 56.4 & 58.3 & 65.6 & 58.2 & 54.5 & 56.4 & 56.4 \\
 & Contextual RAG GritLM-7B & 58.3 & 58.3 & 54.6 & 56.4 & 58.2 & 56.4 & 54.5 \\
 & DRAGIN & 41.8 & 40.0 & 47.3 & 41.8  & 47.3  & 43.6  & 43.6  \\
 & R$^{2}$AG & 54.5 & 58.2 & 61.8 & 50.9  & 49.1  & 50.9  & 50.9  \\
 & LongRAG & 50.9 & 54.6 & 60.0 &47.3 & 49.1  &50.9  & 47.3  \\
\midrule
Agent & LUMOS & 43.6 & 47.3 & 49.1 & 49.1  & 47.3  & 47.3  &52.7  \\
 & PaperQA2 & 38.2 & 45.5 & 47.3 & 47.3  & 45.5  & 47.3  & 50.9  \\
 & Agentic-Hybrid-RAG & 54.5 & 58.2 & 60.0 & 52.7&52.7& 50.9 & 50.9  \\
 & SciMaster & 58.2 & 54.5 & 60.0 & 50.9  &50.9  & 49.1 & 50.9  \\
 & \textbf{\agentname (Ours)} & 58.2 & 69.1 & 69.1 & 65.5 & 67.3 & 70.9 & 69.1 \\
\hline
\end{tabular}

}
    \caption{Accuracies (\%) on \benchname, Public Health
    }
    \label{tab:model-performance-ph}

\end{table*}

\begin{table*}[t!]
  \centering
    \small
\setlength{\tabcolsep}{3.5pt}
\resizebox{\textwidth}{!}{
\begin{tabular}{@{}p{0.6cm}lccccccc@{}}
\hline
&\textbf{Baseline} & \textbf{GPT-4o} & \textbf{GPT-4.1} & \textbf{DS-R1} & \textbf{o3}& \textbf{o4-mini} & \textbf{Gemini-2.5 Pro} & \textbf{Llama-3.1-70B} \\
\hline
RAG & Vanilla RAG all-MiniLM-L6-v2 & 60.0 & 58.3 & 60.0 & 58.3 & 61.7 & 55.0 & 60.0 \\
 & Vanilla RAG E5-mistral-7b-instruct & 60.0 & 60.0 & 58.3 & 58.3 & 61.7 & 51.7 & 60.0 \\
 & Vanilla RAG GritLM-7B & 61.7 & 60.0 & 58.3 & 53.3 & 51.7 & 51.7 & 61.7 \\
 & Contextual RAG E5-mistral-7b-instruct & 58.3 & 61.8 & 60.0 & 51.7 & 51.7 & 58.3 & 61.7 \\
 & Contextual RAG GritLM-7B & 58.3 & 58.2 & 60.0 & 48.3 & 56.7 & 58.3 & 60.0 \\
 & DRAGIN & 40.0 & 43.3 & 40.0 & 43.3  & 45.0  & 46.7  & 46.7  \\
 & R$^{2}$AG & 58.3 & 56.7 & 58.3 & 55.0  & 55.0  &53.3  & 60.0  \\
 & LongRAG & 60.0 & 63.3 & 63.3 & 53.3  & 53.3  & 53.3  & 58.3  \\
\midrule
Agent & LUMOS & 46.7 & 51.7 & 46.7 & 53.3  &56.7  & 51.7  & 55.0  \\
 & PaperQA2 & 51.7 & 50.0 & 53.3 & 50.0  & 45.0  & 48.3  & 55.0  \\
 & Agentic-Hybrid-RAG & 58.3 & 58.3 & 58.3 & 56.7  & 55.0  & 55.0  & 60.0  \\
 & SciMaster & 56.7 & 56.7 & 60.0 & 56.7  & 58.3  & 60.0  & 61.7   \\
 & \textbf{\agentname (Ours)} & 63.3 & 70.0 & 70.0 & 70.0 & 69.1 & 70.1 & 65.0 \\
\hline
\end{tabular}
}
\caption{Accuracies (\%) on \benchname, Earth Science
}
\label{tab:model-performance-es}
\end{table*}

\begin{table*}[t!]
  \centering
    \small
\setlength{\tabcolsep}{3.5pt}
\resizebox{\textwidth}{!}{
\begin{tabular}{@{}p{0.6cm}lccccccc@{}}
\hline
&\textbf{Baseline} & \textbf{GPT-4o} & \textbf{GPT-4.1} & \textbf{DS-R1} & \textbf{o3}& \textbf{o4-mini} & \textbf{Gemini-2.5 Pro} & \textbf{Llama-3.1-70B} \\
\hline
RAG & Vanilla RAG all-MiniLM-L6-v2 & 64.6 & 61.5 & 60.0 & 58.5 & 55.4 & 61.5 & 52.3 \\
 & Vanilla RAG E5-mistral-7b-instruct & 60.0 & 58.5 & 61.5 & 60.0 & 60.0 & 60.0 & 55.4 \\
 & Vanilla RAG GritLM-7B & 63.1 & 56.9 & 61.5 & 61.5 & 63.1 & 63.1 & 55.4 \\
 & Contextual RAG E5-mistral-7b-instruct & 63.1 & 64.3 & 60.0 & 61.5 & 56.9 & 63.1 & 52.3 \\
 & Contextual RAG GritLM-7B & 66.2 & 64.3 & 60.0 & 60.0 & 58.4 & 63.1 & 60.0 \\
 & DRAGIN & 47.7 & 50.8 & 52.3 & 46.2  & 44.6  & 44.6 & 46.2  \\
 & R$^{2}$AG & 60.0 & 61.5 & 63.1 & 60.0  & 56.9  & 56.9  & 56.9 \\
 & LongRAG & 67.7 & 69.2 & 70.8 & 63.1  & 66.2  & 63.1  & 61.5  \\
\midrule
Agent & LUMOS & 55.4 & 58.5 & 64.6 & 58.5  &56.9  & 58.5  & 53.8  \\
 & PaperQA2 & 50.8 & 52.3 & 58.5 & 53.8  &47.7  & 52.3  & 52.3  \\
 & Agentic-Hybrid-RAG & 64.6 & 61.5 & 66.2 & 61.5  & 63.1  & 61.5  & 58.5  \\
 & SciMaster & 64.6 & 60.0 & 66.2 & 61.5  & 61.5  & 60.0  & 56.9  \\
 & \textbf{\agentname (Ours)} & 80.0 & 81.5 & 78.5 & 81.5 & 75.4 & 76.9 & 66.2 \\
\hline
\end{tabular}
}
\caption{Accuracies (\%) on \benchname, Engineering
}
\label{tab:model-performance-hf}
\end{table*}

\begin{table*}[t!]
  \centering

\setlength{\tabcolsep}{3.5pt}
\resizebox{\textwidth}{!}{
\begin{tabular}{@{}p{0.6cm}lccccccc@{}} 
\hline
&\textbf{Baseline} & \textbf{GPT-4o} & \textbf{GPT-4.1} & \textbf{DS-R1} & \textbf{o3}& \textbf{o4-mini} & \textbf{Gemini-2.5 Pro} & \textbf{Llama-3.1-70B} \\
\hline
RAG & Vanilla RAG all-MiniLM-L6-v2 & 64.3 & 60.0 & 71.4 & 68.6 & 68.6 & 74.3 & 61.4 \\
 & Vanilla RAG E5-mistral-7b-instruct & 62.9 & 70.0 & 70.0 & 65.7 & 64.3 & 68.6 & 62.9 \\
 & Vanilla RAG GritLM-7B & 64.3 & 70.0 & 71.4 & 68.6 & 64.3 & 64.3 & 67.1 \\
 & Contextual RAG E5-mistral-7b-instruct & 67.1 & 68.6 & 70.0 & 65.7 & 68.6 & 62.9 & 65.7 \\
 & Contextual RAG GritLM-7B & 67.1 & 65.7 & 68.6 & 68.6 & 70.0 & 70.0 & 62.9 \\
 & DRAGIN & 44.3 & 45.7 & 47.1 & 45.7  & 51.4  & 50.0  & 44.3  \\
 & R$^{2}$AG & 65.7 & 64.3 & 65.7 & 60.0  & 58.6 & 60.0  & 57.1  \\
 & LongRAG & 67.1 & 65.7 & 68.6 & 64.3  & 64.3  & 62.9  & 62.9  \\
\midrule
Agent & LUMOS & 52.9 & 57.1 & 65.7 & 61.4  & 64.3  & 60.0  & 61.4  \\
 & PaperQA2 & 48.6 & 45.7 & 58.6 & 58.6  & 55.7  & 57.1  & 58.6  \\
 & Agentic-Hybrid-RAG & 61.4 & 60.0 & 65.7 & 61.4  & 62.9  & 61.4  & 60.0 \\
 & SciMaster & 57.1 & 58.6 & 67.1 & 62.9  & 61.4  & 61.4  &60.0   \\
 & \textbf{\agentname (Ours)} & 72.9 & 80.0 & 75.7 & 72.9 & 75.7 & 82.9 & 71.4 \\
\hline
\end{tabular}
}
\caption{Accuracies (\%) on \benchname, Material Science}
\label{tab:model-performance-mfg}

\end{table*}

\subsection{Statistical Analysis}
\subsubsection{Significance Testing of \agentname{} Performance}
\label{subsec:wilcoxon}

To rigorously assess whether \agentname\ achieves statistically significant performance gains over existing baselines, 
we conducted a one-sided Wilcoxon signed-rank test across all five domains and seven backbones 
($n = 35$ paired samples per baseline). 
For each baseline, accuracies were paired with those of \agentname\ on identical backbones within the same domain.

\paragraph{Hypotheses}
\begin{align*}
H_0 &: \text{Median}(\text{\agentname} - \text{Baseline}) = 0, \\
H_1 &: \text{Median}(\text{\agentname} - \text{Baseline}) > 0.
\end{align*}
Here $H_0$ indicates no improvement of \agentname\ over the baseline, while $H_1$ tests for significantly higher performance, result shown~\ref{tab:wilcoxon_results}.

\begin{table*}[t!]
\centering
\small
\setlength{\tabcolsep}{3pt}
\begin{tabular}{lccc}
\hline
\textbf{Baseline} & \textbf{Median $\Delta$(\%)} & \textbf{$p_{\text{raw}}$} & \textbf{$p_{\text{adj}}$} \\
\hline
Vanilla RAG all-MiniLM-L6-v2 & +11.7 & $2.9\times10^{-11}$ & $3.5\times10^{-10}$ \\
Vanilla RAG E5-mistral-7b-instruct & +10.9 & $2.9\times10^{-11}$ & $3.5\times10^{-10}$ \\
Vanilla RAG GritLM-7B & +11.7 & $2.9\times10^{-11}$ & $3.5\times10^{-10}$ \\
Contextual RAG E5-mistral-7b-instruct & +12.8 & $2.9\times10^{-11}$ & $3.5\times10^{-10}$ \\
Contextual RAG GritLM-7B & +12.9 & $5.8\times10^{-11}$ & $3.5\times10^{-10}$ \\
DRAGIN & +27.3 & $2.9\times10^{-11}$ & $3.5\times10^{-10}$ \\
R$^{2}$AG & +16.8 & $2.9\times10^{-11}$ & $3.5\times10^{-10}$ \\
LongRAG & +12.3 & $2.9\times10^{-11}$ & $3.5\times10^{-10}$ \\
LUMOS & +18.5 & $2.9\times10^{-11}$ & $3.5\times10^{-10}$ \\
PaperQA2 & +21.8 & $2.9\times10^{-11}$ & $3.5\times10^{-10}$ \\
Agentic-Hybrid-RAG & +14.1 & $2.9\times10^{-11}$ & $3.5\times10^{-10}$ \\
SciMaster & +14.6 & $1.8\times10^{-7}$ & $1.8\times10^{-7}$ \\
\hline
\end{tabular}
\caption{Wilcoxon signed-rank test comparing \agentname\ with all baselines across 35 paired samples (5 domains $\times$ 7 backbones). 
All tests are one-sided ($H_1$: \agentname\ $>$ baseline). 
Reported $p$-values are Holm--Bonferroni corrected for multiple comparisons. 
All comparisons remain significant at $p_{\text{adj}} < 0.001$.}
\label{tab:wilcoxon_results}
\end{table*}

\paragraph{Interpretation}
Across all baselines, the median performance improvements range from $+10.9$ to $+27.3$ percentage points. 
The Holm--Bonferroni–adjusted $p$-values ($p_{\text{adj}} < 10^{-6}$) indicate that these gains are highly significant and 
not attributable to random variation. 
Even after correction for multiple comparisons, all $12$ tests reject the null hypothesis $H_0$ at $\alpha=0.05$. 
These findings confirm that \agentname\ consistently and significantly outperforms every baseline across domains and backbones, 
demonstrating robust, model-agnostic performance advantages.

\subsubsection{Statistical Analysis of Paper Length and \agentname Performance}
\label{subsec:paper_length_stats}

\benchname spans a wide range of paper lengths, with an average of 15.2 pages (SD = 9.04), 
a median of 13 pages, a minimum of 6 pages, and a maximum of 49 pages. To examine whether paper length affects agent performance, we build a mixed-effects linear regression model 
with the per-paper accuracy of \agentname{} as the response and paper length as the predictor. 
Since each paper is evaluated using three backbone LLMs (GPT-4o, GPT-4.1, and DeepSeek-R1), 
we treat backbone choice as repeated measurements.

\paragraph{Hypotheses}
\begin{align*}
H_0 &: \beta_{\text{length}} = 0, \\
H_1 &: \beta_{\text{length}} \neq 0.
\end{align*}
Here, $\beta_{\text{length}}$ denotes the regression coefficient associated with paper length. 
$H_0$ indicates that paper length has no effect on \agentname{}'s performance, 
while $H_1$ tests whether paper length has a statistically significant effect.

\paragraph{Interpretation}
The model estimates a slope coefficient of $-0.1444$ for paper length, with $p = 0.463 > 0.05$ 
and a 95\% confidence interval of $(-0.535,\ 0.246)$. 
Although the estimated coefficient is negative, its magnitude is small, and the $p$-value is well above 
the standard significance threshold of $\alpha = 0.05$. 
Therefore, we fail to reject $H_0$, indicating that paper length is not significantly associated with 
\agentname{}'s performance in the current evaluation.

\subsection{Execution Time Report}\label{subsec: run_time}

In this section, we report the computational runtime required to complete a single test instance on average. The runtime of \agentname is influenced by several factors, including the number of reasoning iterations until information sufficiency is achieved, API connection stability, and the response speed of the backbone LLM. Nevertheless, we provide an approximate runtime in seconds, averaged over multiple runs. (See Table\ref{tab:run_time})

Since the number of runs can vary across experiments, we place greater emphasis on the median runtime on the default setting to provide a more representative estimate. \agentname achieves a median runtime of approximately 129 seconds, or about two minutes per paper per question. In comparison, the baseline methods require slightly less than one minute.

Given the nature of the \taskname task, our primary concern is accuracy rather than speed. Nevertheless, the agent remains significantly faster than human-level reading and reasoning over multiple full-text papers.

\begin{table*}[h!]
\centering
\small
\begin{tabular}{p{3.5cm} p{5.5cm} r}
\toprule
\textbf{Method} & \textbf{Step} & \textbf{Time estimation (sec)} \\
\midrule
Evaluation protocol & MCQ mapping & 5 \\
\midrule
\agentname & Format conversion & 36 \\
 & Section Ranking stage & 4 \\
 & Iterative Reading stage (1 iter) & 42 \\
 & Median (Default setting) & 129 \\
 & Average (Default setting) & 259 \\
 \midrule
Vanilla RAG & Plaintext conversion & 2 \\
 & Short answer generation & 35 \\
 \midrule
 DRAGIN & Plaintext conversion & 2 \\
 & Paragraph-Level Corpus Construction & 4 \\
 & Chain-of-Thought Construction(each one) & 23 \\
 & Step-by-step Reasoning & 55 \\
 \midrule
R\textsuperscript{2}AG & Plaintext conversion & 2 \\
 & Retriever Embedding & 2 \\
 & Ranking & 5 \\
 & Reranking & 10 \\
 & Answer Generation with Retrieval Guidance &30  \\
\bottomrule
Lumos & Plaintext conversion & 2 \\
 & Planning for a single Subgoal & 6 \\
 & Converting Subgoal into a Retrieval Action & 2 \\
 & Execution &2 \\
 \midrule
\end{tabular}
\caption{Estimated runtime (in seconds) for \agentname and baselines. Running with GPT-4o. Values are averaged over multiple runs per test instance.}
\label{tab:run_time}
\end{table*}
 
\section{Ablation Studies Detail}
\label{sec:Ablation studies detail}

\subsection{Failure Case Study}\label{subsec: Failure Case}
\subsubsection{Hallucination Failure Case from Missing Information Sufficiency Check}\label{subsubsec: Hallucination Failure Case}

As an extension of our ablation study, we examine a specific failure mode caused by removing the information sufficiency check from the Iterative Reading stage. This case highlights the critical role of sufficiency verification in preventing answer hallucination.

We analyze a representative question from the \textit{SERS} dataset based on the paper by Rathnayake et al.~\cite{rathnayake_machine_2024}:
The correct section is successfully retrieved by section ranking stag: \textit{Experimental - Surface entrapping of AGP and SERS measurements}. The following detail is extracted:
\begin{quote}
``The measurements were performed on a ProRaman-L spectrometer with \textbf{785~nm excitation wavelength} using a $40\times$ objective, under the following conditions: power density of approximately $40~\upmu\mathrm{W}/\upmu\mathrm{m}^2$, accumulation time of $40~\mathrm{s}$, averaging over 3 spectra, and a spectral range of $400$--$3000~\mathrm{cm}^{-1}$.'' 
\end{quote}

This evidence directly supports the correct answer: \textbf{785~nm}.

In contrast, when the information sufficiency check is disabled, the agent outputs a fully hallucinated answer:

\begin{quote}
``The excitation laser wavelength used for SERS measurements is confirmed to be \textbf{532~nm}. This is supported by the direct statement: ‘The excitation wavelength for the SERS measurements is 532~nm’ found in the text (Chunk 3, Section\_title: Methodology).'' 
\end{quote}

In this case, the cited sentence, section title, and numerical value are entirely fabricated and do not exist in the original document. The agent not only provides an incorrect answer but does so with unjustified confidence and fabricated attribution.

This failure mode demonstrates how critical the information sufficiency check is for ensuring factual grounding. Its absence increases the risk of hallucination, where the system generates plausible-sounding yet entirely unsupported responses.

\subsubsection{Contextual Relationship Blindness from Domain-Specific Knowledge Gaps}\label{subsubsec: Contextual Relationship Blindness}

Another failure pattern not resolved simply by using a more advanced LLM is \textbf{Contextual Relationship Blindness}. We believe this is partly due to gaps in domain knowledge: during the \textit{Section Detail Extraction} phase, the model fails to establish the correct relationship between the question and the corresponding sections in a given iteration. As a result, it may overlook crucial information, leading to deficiencies during the information sufficiency check stage and causing the system to loop to later sections, ultimately selecting ``None of the Above.''

For example, in our Earth Science dataset (specifically, the subfield of urban remote sensing), the main study targets are urban landscapes or buildings. Consider question \#8 from the batch:

\begin{quote}
Are auxiliary features used beyond raw spectral bands?
\end{quote}

In remote sensing, it is well understood that the main features are spectral bands and radar bands. Auxiliary features are additional derived features, such as vegetation indices (NDVI, LAI, FAPAR), land surface temperature, albedo, emissivity, topographic information (e.g., DEM), or other external variables. However, when the LLM does not pick up the meaning or understand the term, even if the paragraph includes terms like NDVI, the relationship may still be overlooked. Cases like these highlighting \benchname remain challenging. To improve the usability of the dataset for researchers who are not familiar with cross-domain terminology, we have annotated these terms in the benchmark; however, these annotations are \textbf{not} provided as inputs to the agent.

\section{Robustness of \agentname to Mapping Model and Input Variability Detail}
\label{sec:OOD studies detail}

\subsection{Section Heading Rework}\label{appx:bad_title}

In Section~\ref{subsec:OOD}, we investigated how \agentname performs when provided with subpar or non-standard section headings in scientific literature. Specifically, we examine whether the agent can maintain accuracy even when the section structure is rewritten in unfamiliar or distorted styles.

To conduct this study, we extracted all section headings from the paper by Thrift et al.~\cite{thrift_quantification_2019}, listed in the first column of Table~\ref{tab: Section Heading Rework}. We then created three alternate versions of these headings: a simplified "Beginner Style", a distorted "Highly Noisy Version", and a stylized "Shakespearean Style", each shown in the subsequent columns.

For each variant, we replaced the original section titles in the document with the newly generated versions. \agentname was then applied to the modified texts using GPT-4o as the backbone LLM and GPT-4.1 as the mapping model, under default settings. 
\onecolumn
{\small
\begin{longtable}{%
  >{\raggedright\arraybackslash}p{0.22\textwidth} |
  >{\raggedright\arraybackslash}p{0.22\textwidth} |
  >{\raggedright\arraybackslash}p{0.22\textwidth} |
  >{\raggedright\arraybackslash}p{0.22\textwidth}
}
\toprule
\textbf{Original Title} & \textbf{Beginner Style} & \textbf{Highly Noisy Version} & \textbf{Shakespearean Style} \\
\midrule
\endfirsthead
\toprule
\textbf{Original Title} & \textbf{Beginner Style} & \textbf{Highly Noisy Version} & \textbf{Shakespearean Style} \\
\midrule
\endhead

A R T I C L E  I N F O & Stuff About This Paper & ZXCV ARTICLE BLAH & Prologue of This Learned Work \\
A B S T R A C T & What This Thing’s About & QWERT ABSTRACT THING & Abstract of Most Worthy Endeavor \\
1. Introduction & 1. Starting Out & 1. asdjkh Introduction blah & 1. An Overture to the Matter at Hand \\
2. Materials and methods & 2. Stuff We Used and Did & 2. jkjhas Materials and doings & 2. Of Materials Gathered and Deeds Undertaken \\
2.1. Bacteria and sample preparation & 2.1. Germs and How We Got Them & 2.1. ajskdfh Bacteria things and getting stuff & 2.1. Of Bacterium’s Harvest and Preparations Befitting Study \\
2.2. AFM imaging & 2.2. Taking Tiny Pics & 2.2. kjashdf AFM looking-at-things & 2.2. Wherein the Art of Atomic Force Is Employed to Gaze Upon the Minuscule \\
2.3. SERS measurements & 2.3. Laser Stuff on the Germs & 2.3. asdkljfh SERS zappy laser stuff & 2.3. Of Golden Gleams and SERS—The Spectral Oracle \\
2.4. Machine learning model development & 2.4. Computer Things We Tried & 2.4. zmxnv Machine learning magic box & 2.4. Wherein the Machine, Apt in Learning, Is Taught to Discern \\
3. Results & 3. What Happened & 3. kjsdhf Results what happened & 3. Revelations and Findings Most Curious \\
3.1. AFM analysis of microbial shape and dimension & 3.1. Shapes of the Germs & 3.1. qwekjh Germ shapes and sizes & 3.1. Of Form and Measure—The Shape of the Microscopic Realm \\
3.2. SERS signature of periodontal pathogens & 3.2. Shiny Lights from Germs & 3.2. kjasdh SERS noise from germs & 3.2. Upon the Light’s Whisper: Signatures of the Mouth’s Hidden Foes \\
3.3. ML-enabled identification of periodontal pathogens & 3.3. Using Computer to Guess Germs & 3.3. hgjdksh Computer guessing germ names & 3.3. The Thinking Engine’s Triumph in Unmasking the Pathogen \\
4. Discussion & 4. What We Think It Means & 4. lkjdfh What we think (maybe?) & 4. A Discourse Upon the Meaning of These Happenings \\
5. Conclusion & 5. Wrapping It Up & 5. THE END (idk lol) & 5. A Summation, As the Curtain Draws Near \\
CRediT authorship contribution statement & Who Did What & blorpflorp CRediT who did what maybe & Of Quills and Labors: A Testament to Those Who Contributed \\
Declaration of competing interest & We Don’t Got Any Fights About This & nope-no-fightz Declaration or whatever & Of Conflicts, None Declared nor Concealed \\
Data availability & Where the Stuff Is & canhasdata Data place?? & Where Might the Curious Find the Data? \\
Acknowledgement & Thanks and Shoutouts & thxbye Acknowlegmints & Gratitude, in Verse and Spirit \\
Appendix A. Supplementary data & Extra Stuff We Didn’t Put Up There & Appendix A. More junk we had & Appendix the First: Scrolls of Supplement Yet Untold \\
References & Books and Papers We Looked At & Books n Stuff We Readed (aka refs) & Tomes and Records Consulted—A Catalog of the Wise \\
\bottomrule

\caption{Rewriting original section headings from Thrift et al.~\cite{thrift_quantification_2019} into three alternative styles—Beginner Style, Highly Noisy Version, and Shakespearean Style—to evaluate the robustness of \agentname to variation in document structure.} \label{tab: Section Heading Rework} \\
\end{longtable}}
\twocolumn

\subsection{Section Heading Rework - Highly Noisy Version Reasoning Example}\label{appx:bad_title_exp}

This is a detailed reasoning example from the SERS dataset Thrift et al.~\cite{thrift_quantification_2019}, using the second section title rewrite variation, labeled as the Highly Noisy Version. The rewritten section headings are provided in Table~\ref{tab: Section Heading Rework}.

With paper main title provided alongside with the research question for this case is: What are the material and structural properties, or the morphology, of the SERS substrates used?

After the Hierarchy Preservation step, the processed section headings are as follows: 
\begin{quote}
ZZXCV ARTICLE BLAH; QWERT ABSTRACT THING; asdjkh Introduction blah; 2. jkjhas Materials and doings - 2.1. ajskdfh Bacteria things and getting stuff; 2. jkjhas Materials and doings - 2.2. kjashdf AFM looking-at-things; 2. jkjhas Materials and doings - 2.3. asdkljfh SERS zappy laser stuff; 2. jkjhas Materials and doings - 2.4. zmxnv Machine learning magic box; 3. R kjsdhf Results what happened - 3.1. qwekjh Germ shapes and sizes; 3. R kjsdhf Results what happened - 3.2. kjasdh SERS noise from germs; 3. R kjsdhf Results what happened - 3.3. Mhgjdksh Computer guessing germ names; 4. lkjdfh What we think (maybe?); 5. THE END (idk lol); blorpflorp CRediT who did what maybe; nope-no-fightz Declaration or whatever; canhasdata Data place??; thxbye Acknowlegmints; more junk we had; Books n Stuff We Readed (aka refs)
\end{quote}

In semantic section ranking, we derived the reasoning step:
\begin{quote}
1. 2. jkjhas Materials and doings - 2.3. asdkljfh SERS zappy laser stuff — This section specifically mentions "SERS" in its title, indicating it likely contains detailed information about the SERS substrates, including their materials and structure or morphology.

2. 2. jkjhas Materials and doings - 2.1. ajskdfh Bacteria things and getting stuff — As part of the "Materials and doings" main section, this subsection may discuss materials related to the experimental setup, possibly including the SERS substrates.

3. 2. jkjhas Materials and doings - 2.2. kjashdf AFM looking-at-things — This subsection likely involves materials characterization using AFM, which may include details about substrate morphology or materials.

4. asdjkh Introduction blah — The introduction often contains a summary of the materials and methods used, potentially offering an overview of the SERS substrate materials and structure.

...

\end{quote}

This ultimately leads to correct section allocation, enabling detailed information gathering and passing the information sufficiency check without difficulty. It also results in accurate final mapping to the correct multiple-choice answer. This example highlights the model’s ability to extract relevant information from scrambled headings by reasoning through their underlying intent, even in the presence of noise during section ranking.

\input{prompt}

\section{Baseline Evaluation Details}
\label{sec:Baseline Evaluation}
\subsection{Vanilla RAG}

Vanilla RAG refers to the original implementation proposed by Lewis et al.~\cite{10.5555/3495724.3496517}. In our setup, we first convert the PDF versions of papers into plain text. Subsequently, we tokenize the text into fixed-length chunks of 500 tokens with an overlap of 50 tokens between consecutive chunks. We select 500 tokens as it provides a balance between semantic coherence and memory efficiency, allowing enough context to be preserved within each chunk. The 50-token overlap ensures that important information occurring at chunk boundaries is not lost, improving retrieval continuity and robustness.

For embedding generation and similarity computation, we use the \texttt{all-MiniLM-L6-v2304} model to encode each chunk. During retrieval, we adopt a top-$k$ strategy, defaulting to the top 3 chunks most similar to the given research question, based on cosine similarity scores. This top-$k$ selection aligns with our evaluation protocol, where these top 3 chunks are passed into the generation module to produce short-form answers.

For final answer generation, we evaluate selected backbone LLMs under identical input conditions to ensure a fair comparison. Each model generates an answer conditioned on the retrieved top-3 chunks and the research question.

\subsection{Contexual RAG}\label{subsec:contexualrag}

To enhance retrieval precision beyond conventional dense embedding approaches, we implemented a Contextual Retrieval-Augmented Generation (Contextual RAG) framework~\cite{anthropic2024contextual}. Unlike vanilla RAG, which embeds document chunks in isolation, Contextual RAG explicitly augments each chunk with a succinct, model-generated context situating it within the broader document structure. This implementation effectively bridges semantic and structural understanding, allowing the retrieval component to capture both what is being discussed and where it fits contextually within the paper. This contextual enrichment helps disambiguate semantically similar fragments, improving the embedding model’s ability to capture fine-grained relationships between queries and relevant content.

Our pipeline first extracts raw text given literature, followed by cleaning and token-level chunking with a 500-token window and 50-token overlap to preserve local continuity. For each chunk, a lightweight contextualization module built on GPT-4o generates concise summaries that describe how the chunk fits within the overall document; prompt used see~\ref{prompt:contexual RAG}. These augmented chunks are then encoded using models—either E5-Mistral-7B-Instruct or GritLM-7B— for efficient batch inference. Query and chunk embeddings are computed with cosine similarity to identify the top-k most relevant text segments per question, enabling high-precision retrieval aligned with the document’s narrative structure.

\subsection{DRAGIN}
We adopt \textit{DRAGIN} (Dynamic Retrieval Augmented Generation based on the real-time Information Needs of LLMs)\cite{suDRAGINDynamicRetrieval2024} as the baseline framework for dynamic retrieval-augmented generation. \textit{DRAGIN} explicitly determines both \textit{when to retrieve} and \textit{what to retrieve} based on the model’s internal uncertainty and token-level salience signals, enabling fine-grained control over evidence acquisition during generation.

To adapt \textit{DRAGIN} to our customized scenario involving a small-scale paragraph-level corpus and chain-of-thought-style few-shot prompting, we introduce the following modifications:
\begin{itemize}
\item \textbf{Paragraph-Level Corpus Construction.} We replace the default Wikipedia corpus (\texttt{psgs\_w100.tsv}) used in the official \textit{DRAGIN} release with a custom paragraph-level corpus (\texttt{DRAGIN\_paragraphs.tsv}) tailored to our domain. Each entry consists of a unique paragraph ID, a document title, and the paragraph text. This design enables the retrieval module to operate at a finer granularity, which is essential for scenarios involving a small collection of five domain-specific documents.

\item \textbf{Few-Shot Prompting with Summary-Driven CoT Chains.} We generate ten chain-of-thought (CoT) reasoning exemplars from the five domain-specific documents. For each test instance, \agentname summarizes the top five retrieved paragraphs to construct a condensed context representation. These summaries serve as inputs to guide CoT generation and are incorporated into the \texttt{demo} field during the \texttt{inference()} stage. For example:
\begin{quote}
“The spatial extent of the study area is approximately 966 km², derived from the Residential category's area (398 km²) and its proportion (41.20\%) of the total impervious surface...”
\end{quote}

This procedure ensures that the few-shot prompt incorporates domain-relevant knowledge in a structured format, strengthening the alignment between the model's reasoning path and the underlying evidence.

\item \textbf{Dynamic Retrieval at Inference Time.} During inference, \textit{DRAGIN} actively monitors for hallucination-prone segments and triggers retrieval of relevant evidence in real time. Retrieved paragraphs are incrementally integrated into the generation context, enabling iterative refinement of the model’s output. Unlike short-form QA systems, \textit{DRAGIN} produces complete chain-of-thought reasoning sequences, along with auxiliary metadata such as retrieval count and token usage for diagnostic analysis.
\end{itemize}

Following CoT generation, we initiate a final synthesis stage. The full reasoning trace, rather than the retrieved paragraphs themselves, is incorporated into a standardized prompt alongside the original question. This prompt is then passed to an external language model (e.g., GPT-4-turbo, GPT-4o, or DeepSeek-Chat), which is instructed to summarize the reasoning and produce a concise final answer. This decoupled architecture ensures interpretability during intermediate steps while maintaining fluency and coherence in the final output.

\subsection{R$^2$AG}
We adopt \textit{R$^2$AG} (Reranking-augmented Retrieval-Augmented Generation) as one of our baseline\cite{ye_r2ag_2024}. \textit{R$^2$AG} enhances generation quality by incorporating a reranking module that explicitly guides evidence selection, thus mitigating retrieval noise and improving relevance alignment in the final output.

To implement the \textit{R$^2$AG} pipeline in our setting, we follow the procedures below:

\textbf{Retriever Fine-Tuning and Dense Indexing.}
We fine-tune a Sentence-BERT retriever using positive and negative question-paragraph pairs. Specifically, we leverage the DuReader benchmark, which provides human-annotated relevance labels for question-passage pairs, ensuring high-quality supervision during training. Positive samples are passages annotated as relevant to the given question, while negatives are randomly selected or chosen as hard negatives from the same context. After training, the retriever encodes all corpus paragraphs into dense embeddings, which are then indexed using FAISS (\texttt{IndexFlatIP}) for efficient top-$k$ similarity search during retrieval.

\textbf{Construction of\textit{R$^2$AG} Training Samples.}
Each training instance comprises a question, reference answer, $k$ retrieved paragraph embeddings, and a binary relevance label for each paragraph. Labels are heuristically assigned based on similarity with the gold answer using token overlap or ROUGE. These samples are used to jointly supervise both generation and reranking.

\textbf{Relevance-Aware Generation via RFormer.}
During training, paragraph embeddings are processed by \textit{R-Former}, a lightweight transformer encoder that outputs a relevance-guided vector. This vector is projected to match the hidden dimension of a frozen LLM (Here we used LLaMA) and injected into the prompt via a placeholder token. The model jointly optimizes generation loss (over the answer) and binary classification loss (over paragraph relevance).

\textbf{Inference}
At inference time, \textit{R$^2$AG} retrieves top-$k$ passages, processes them via RFormer, and injects the reranking-aware signal into the prompt before decoding with the LLM.

\subsection{LongRAG}
We adopt \textit{LongRAG}~\cite{jiang2024longragenhancingretrieval} as one of our baselines.
\textit{LongRAG} enhances retrieval-augmented reasoning through a dual-perspective design consisting of a \textit{Hybrid Retriever} for coarse-to-fine semantic recall, an \textit{LLM-augmented Information Extractor} for global context recovery, and a \textit{CoT-guided Filter} for relevance refinement.
To adapt \textit{LongRAG} to our scientific literature corpus and domain-specific QA task, we apply several modifications.

First, each paper is segmented into semantically coherent chunks of approximately 200 words, using sentence boundaries as the minimal division unit.
Each chunk is encoded with the \texttt{intfloat/multilingual-e5-large} dual encoder and stored in a FAISS index for dense retrieval.
In the fine-grained re-ranking stage, the cross-encoder \texttt{nreimers/mmarco-mMiniLMv2-L12-H384-v1} is used with default settings \texttt{(chunk\_size = 200, top-k = 7)}.

Second, the retrieved chunks are mapped back to their original paragraphs to restore paragraph-level semantic continuity.
When multiple chunks correspond to the same paragraph, only the most relevant one is retained.
The resulting paragraphs are passed to an LLM-based extractor, which generates a global contextual representation capturing structural and semantic relationships across the paper.

Third, we retain LongRAG’s two-stage CoT-guided filtering mechanism to eliminate redundant or irrelevant content.
The model produces intermediate reasoning traces to identify evidence-bearing segments and synthesizes an intermediate answer by combining globally summarized and filtered information.
In our adaptation, GPT-3.5-Turbo-16k serves as the backbone model for extraction, filtering, and generation.

Finally, a more capable external LLM—such as GPT-4o, GPT-4.1, DeepSeek-R1, Gemini 2.5 Pro, o3, o4-mini, or Llama-3.1-70B-Instruct-Turbo—is employed to refine intermediate outputs and produce the final synthesized answers.
Accuracy is reported as the primary evaluation metric, consistent with our multiple-choice benchmark protocol.
The models used in each stage are summarized in Table~\ref{tab:longrag_models}.
\begin{table*}[t!]
\centering
\small

\begin{tabular}{p{4.5cm} p{10cm}}
\toprule
\textbf{Stage} & \textbf{Model / Tool} \\
\midrule
Dual-Encoder Retrieval & \texttt{intfloat/multilingual-e5-large} \\
Cross-Encoder Re-ranking & \texttt{nreimers/mmarco-mMiniLMv2-L12-H384-v1} \\
Information Extraction (Global) & \texttt{GPT-3.5-Turbo-16k} \\
CoT-guided Filtering (Local) & \texttt{GPT-3.5-Turbo-16k} \\
Answer Generation (Intermediate) & \texttt{GPT-3.5-Turbo-16k} \\
Final Answer  & External LLMs: \texttt{GPT-4o}, \texttt{GPT-4.1}, \texttt{DeepSeek-R1}, \texttt{Gemini 2.5 Pro}, \texttt{o3}, \texttt{o4-mini}, \texttt{Llama-3.1-70B-Instruct-Turbo} \\
\bottomrule
\end{tabular}
\caption{Models used in each stage of the \textit{LongRAG} baseline.}
\label{tab:longrag_models}
\end{table*}

\subsection{Agent LUMOS}\label{subsec:LUMOS detail}
This section details our customized version of the LUMOS iterative framework\cite{yinAgentLumosUnified2024}, adapted specifically for the multiple-choice question (MCQ) task. Unlike the original LUMOS framework designed for open-domain QA, our version retains its iterative structure with separate \textbf{Planning}, \textbf{Grounding}, and \textbf{Execution} modules. At each step, the system generates subgoals, formulates paragraph retrieval actions, retrieves relevant evidence, and accumulates results. Once enough evidence is collected, we use an external LLM (GPT-4o, GPT-4-turbo, or DeepSeek-Chat) to synthesize the final answer based on all retrieved content.

Our method retains LUMOS's modular design, consisting of a planning module, grounding module, and execution module. The models used in each stage are summarized in Table~\ref{tab:agent-module-models}:

\begin{table*}[t!]
  \centering
  \small 
  \begin{tabular}{l p{7cm} p{4cm}}
    \toprule
    \textbf{Module} & \textbf{Role} & \textbf{Model Used} \\
    \midrule
    Planning & Generate reasoning subgoals & \texttt{LLaMA 2–7B} \\
    Grounding & Convert subgoals into retrieval actions & \texttt{/} \\
    Execution & Execute retrieval over provided context & \texttt{dpr-reader-multiset-base} \\
    Final Answer & Synthesize final answer using retrieved content & External LLM (GPT-4-Turbo, GPT-4o, or DeepSeek-Chat) \\
    \bottomrule
  \end{tabular}
\caption{Models used in each stage of the \textit{LUMOS} baseline.}
  \label{tab:agent-module-models}
\end{table*}

The baseline system adopts an iterative multi-stage reasoning framework composed of four core modules: planning, grounding, execution, and final synthesis.

The process begins by initializing an intermediate result dictionary with the original question and an initial context. In each iteration, the planning module generates a reasoning subgoal based on the question, accumulated results, and previous actions. This is achieved using a pretrained model, \texttt{lumos\_complex\_qa\_plan\_iterative}, which is built on LLaMA 2--7B. The iteration process terminates either when the planning module outputs an explicit stop signal (e.g., ``Yes, I will stop planning.'') or when a fixed maximum of 5 steps is reached.

Once a subgoal is generated, the system proceeds to the grounding stage, where the subgoal is translated into a structured paragraph retrieval action conditioned on the current reasoning context.

The execution module is responsible for retrieving the most relevant paragraph. It segments the context into overlapping chunks and selects the top match based on semantic similarity using a custom retriever. To evaluate the effect of evidence granularity, we test retrieval based on the top-5, top-7, and top-9 most relevant paragraphs.

After sufficient evidence has been collected through iterations, the system enters the final synthesis stage. All retrieved content is compiled into a standardized prompt, including the original question and numbered evidence chunks. This prompt is then passed to an external large language model---such as GPT-4-turbo, GPT-4o, or DeepSeek-Chat---to generate the final answer. The model is instructed to synthesize an answer \emph{strictly based on the retrieved content}, ensuring transparency and faithfulness.

This modular design supports step-wise and interpretable reasoning, separates control logic from retrieval, and uses the LLM exclusively in the final answer synthesis stage.

\begin{table*}[t!]
  \centering
  \small
  \setlength{\tabcolsep}{3.5pt}
  \begin{tabular}{>{\raggedright\arraybackslash}p{3.2cm} *{13}{c}}
    \toprule
    & & \multicolumn{4}{c}{GPT-4o} & \multicolumn{4}{c}{GPT-4.1} & \multicolumn{4}{c}{DeepSeek-R1} \\
    \cmidrule(r){3-6} \cmidrule(r){7-10} \cmidrule(r){11-14}
    & & Phys & PH & ES & Avg. & Phys & PH & ES & Avg. & Phys & PH & ES & Avg. \\
    \midrule
    LUMOS (3 chunk)             & & 43.1 & 36.4 & 41.7 & 40.4 & 41.5 & 41.8 & 40.0 & 41.1 & 38.5 & 40.0 & 38.3 & 38.9 \\
    LUMOS (5 chunk\textsuperscript{†})  & & 52.3 & 43.6 & 46.7 & 47.5 & 50.8 & 47.3 & 51.7 & 49.9 & 50.8 & 49.1 & 46.7 & 48.9 \\
    LUMOS (7 chunk)             & & 58.5 & 49.1 & 51.7 & 53.1 & 56.9 & 52.7 & 55.0 & 54.9 & 55.4 & 56.4 & 51.7 & 54.5 \\
    LUMOS (9 chunk)             & & 61.5 & 52.7 & 53.3 & 55.8 & 60.0 & 56.4 & 56.7 & 57.7 & 58.5 & 61.8 & 55.0 & 58.4 \\
    \agentname (ours)           & & \textbf{75.4} & \textbf{58.2} & \textbf{63.3} & \textbf{65.6} & \textbf{78.5} & \textbf{69.1} & \textbf{70.0} & \textbf{72.5} & \textbf{72.3} & \textbf{67.3} & \textbf{63.3} & \textbf{67.6} \\
    \bottomrule
  \end{tabular}
  \caption{Accuracy (in percentage) of LUMOS across the three domains in \benchname—SERS in chemistry physics (Phys), \textit{infectious‐disease modeling} in public health (PH), and \textit{remote sensing} in earth science (ES)—under varying chunk configurations, with GPT-4o, GPT-4.1, and DeepSeek-R1 backbones.
  LUMOS achieves consistent performance gains as chunk size increases, peaking at 55.8\%, 57.7\%, and 58.4\% in the Phys, PH, and ES domains, respectively. The default 5-chunk setup strikes a balance between accuracy and efficiency across all models.}
  \label{tab:LUMOS}
  \begin{tablenotes}
    \small
    \item \textbf{Note.} LUMOS with 5 chunks\textsuperscript{†} is the default setting.
  \end{tablenotes}
\end{table*}

\subsection{PaperQA2}

We reproduce and adapt the \textit{PaperQA2} framework for our controlled setting involving a small fixed corpus of input PDFs\cite{PaperQA2}, rather than a large open-domain scientific literature database. The original PaperQA2 pipeline includes dynamic \textit{Paper Search} and \textit{Citation Traversal} modules, which rely on inter-paper citation graphs and large-scale retrieval. Since our corpus consists of a limited number of static papers without citation links, these modules are disabled in our adaptation. Instead, we preserve the core reasoning-based retrieval and summarization components of PaperQA2, ensuring a faithful yet context-appropriate baseline implementation.

Each question–document pair is processed through a streamlined multi-stage pipeline consisting of: 
(1) \textbf{Document Parsing}, where each input PDF is converted to structured text using \texttt{Grobid}; 
(2) \textbf{Chunking \& Indexing}, where parsed text is segmented into semantically coherent units and indexed with \texttt{FAISS}; 
(3) \textbf{Dense Retrieval}, using \texttt{intfloat/multilingual-e5-large} to identify the most relevant chunks to the input query; 
(4) \textbf{RCS (Contextual Summarization)}, where \texttt{GPT-4o} performs reasoning-based summarization and relevance scoring to produce high-quality evidence summaries; and 
(5) \textbf{Answer Generation}, where the top-ranked evidence summaries are consolidated and passed to external LLMs (\texttt{GPT-4o}, \texttt{GPT-4.1}, \texttt{DeepSeek-R1}, \texttt{Gemini 2.5 Pro}, \texttt{o3}, \texttt{o4-mini}, \texttt{Llama-3.1-70B-Instruct-Turbo}) for final response synthesis. 
All models operate with temperature fixed at 0.0 to ensure deterministic factual outputs.

The dense retriever (\texttt{intfloat/multilingual-e5-large}) encodes both query and chunk representations under cosine similarity. For each question, the top-$k$ = 30 retrieved chunks are summarized by the RCS stage. The final answer generation step utilizes one of the listed external LLMs, which directly synthesize concise, evidence-grounded answers. The overall model configuration and stage-wise mapping are summarized in Table~\ref{tab:paperqa2_models}.

\begin{table*}[t!]
\centering
\small

\begin{tabular}{
    >{\raggedright\arraybackslash}p{4.2cm} 
    >{\raggedright\arraybackslash}p{10cm}
}
\toprule
\textbf{Stage} & \textbf{Model / Tool} \\
\midrule
Document Parsing & \texttt{Grobid} (structure-aware scientific text parser) \\
Dense Retrieval & \texttt{intfloat/multilingual-e5-large} (dual encoder for chunk retrieval) \\
RCS (Contextual Summarization) & \texttt{GPT-4o} (reasoning-based summarization and relevance scoring) \\
Final Answer & External LLMs: \texttt{GPT-4o}, \texttt{GPT-4.1}, \texttt{DeepSeek-R1}, \texttt{Gemini 2.5 Pro}, \texttt{o3}, \texttt{o4-mini}, \texttt{Llama-3.1-70B-Instruct-Turbo} \\
\bottomrule
\end{tabular}
\caption{Models used in each stage of the \textit{PaperQA2} baseline.}
\label{tab:paperqa2_models}
\end{table*}

\subsection{Agentic-Hybrid-Rag }
\label{subsec:hybridrag_local}
This section presents the implementation and evaluation setup of the \textit{Agentic-Hybrid-Rag} baseline\cite{AgenticHybridRag}, adapted to our experimental environment featuring a small, offline corpus consisting of a few domain-specific PDF documents. Since the original \textit{Agentic-Hybrid-Rag} framework relies on large-scale online bibliographic sources (e.g., PubMed, ArXiv, Google Scholar) and a server-based Neo4j knowledge graph, we develop a lightweight variant suitable for a fully local, closed-data scenario.

The customized version retains \textit{Agentic-Hybrid-Rag}’s dual-retrieval architecture—combining \textbf{graph-based retrieval} and \textbf{vector-based retrieval}—but replaces all cloud-dependent components with locally executable modules. The system is organized into three main stages: \textbf{Document Preprocessing}, \textbf{Dual Retrieval}, and \textbf{Final Answer Synthesis}. The models and resources used are summarized in Table~\ref{tab:hybridrag_local_models}.

\begin{table*}[t!]
  \centering
  
  \small
  \begin{tabular}{l p{12cm} }
    \toprule
    \textbf{Stage}  & \textbf{Implementation / Model Used} \\
    \midrule
    Graph-based Retrieval   &  \texttt{NetworkX}( Query structured metadata relations) \\
    Vector-based Retrieval  & \texttt{all-MiniLM-L6-v2} + BM25 + FAISS (Retrieve semantically similar passages using hybrid sparse–dense retrieval)\\
    Final Answer & External LLMs: \texttt{GPT-4o}, \texttt{GPT-4.1}, \texttt{DeepSeek-R1}, \texttt{Gemini 2.5 Pro}, \texttt{o3}, \texttt{o4-mini}, \texttt{Llama-3.1-70B-Instruct-Turbo}  (Synthesize a concise, evidence-grounded response based on all retrieved information)\\
    \bottomrule
  \end{tabular}
  \caption{Models used in each stage of the \textit{Agentic-Hybrid-Rag} baseline.}
  \label{tab:hybridrag_local_models}
\end{table*}

Each PDF in the corpus is first converted to structured text using \texttt{pdfminer.six}. Section headers and paragraphs are segmented based on typographic cues, and both the extracted text and associated metadata (title, author, year, and keywords) are serialized into JSON format. The resulting dataset thus supports both symbolic (graph) and semantic (vector) indexing without external dependencies.

To emulate HybridRAG’s structured reasoning component without a full Neo4j server, we construct an in-memory knowledge graph using \texttt{NetworkX}. Nodes represent papers, authors, and keywords, while edges encode relations such as \texttt{written\_by} and \texttt{has\_keyword}. Given a question involving bibliographic relations (e.g., “Which study by Author X discusses…?”), the agent issues rule-based lookups that traverse this lightweight graph and retrieve the corresponding document set.

For semantic retrieval, all document sections are embedded using the \texttt{all-MiniLM-L6-v2} sentence transformer and indexed in FAISS. During inference, queries are encoded into the same embedding space, and the top-$k$ passages are selected based on cosine similarity. To approximate HybridRAG’s hybrid search strategy, BM25 lexical scores are linearly combined with dense similarity scores before reranking. This design ensures that both surface-level and semantic signals are leveraged despite the small data scale.

The retrieved passages from both retrieval modes are merged into a unified context, ordered by relevance, and passed to an answer generator. For fairness across baselines, we use the same instruction-tuned model (\texttt{Mistral-7B-Instruct}) for all answer synthesis tasks. The model is explicitly prompted to reason \emph{only over the retrieved evidence} to avoid hallucination.

The baseline is evaluated on the same set of expert-curated questions used for \agentname. For each query, the router decides between graph or vector retrieval depending on the question type. We report accuracy using the LLM-grounded multiple-choice evaluation metric described in Section~\ref{subsec:bench_evaluation}. This adaptation allows HybridRAG to operate fully offline while preserving its dual-retrieval reasoning capability, thus serving as a reproducible and interpretable baseline for comparison with \agentname.

\subsection{SciMaster}
We adopt \textit{SciMaster}~\cite{chai2025scimaster} as the baseline framework for tool-integrated agentic reasoning. \textit{SciMaster} enables a reasoning model to dynamically interact with external environments through Python-based code execution, combining tool-augmented reasoning with a multi-stage inference-time workflow involving \textit{Solver}, \textit{Critic}, \textit{Rewriter}, and \textit{Selector} roles. To adapt \textit{SciMaster} to our task setting involving domain-specific scientific literature and structured information extraction, we introduce several modifications.

First, we perform \textbf{Domain-Specific Query Adaptation}. Each research question from our benchmark is reformulated into SciMaster's standard input template, consisting of a user query followed by a reasoning block enclosed within \texttt{<think>} and \texttt{</think>} tags. This ensures that the baseline reasoning process remains consistent with the original workflow while aligning the query semantics with scientific reading tasks. The same backbone model, \textit{DeepSeek-R1-0528}, is employed with a temperature of 0.6 and a context length of 64k tokens, identical to the configuration reported in the original paper.

We replace the original general-purpose modules \texttt{web\_search} and \texttt{web\_parse} with a customized local retrieval interface, redirecting all SciMaster retrieval requests to our curated corpus of scientific papers rather than the open web. This design enables controlled access to factual content and ensures a fair comparison with our task-specific agent operating on the same literature sources. The retrieved text snippets are returned as simulated execution results enclosed within \texttt{\textless execution\_results\textgreater} tags, thereby fully preserving the original reasoning--action--feedback loop.

Following the canonical design, the inference pipeline consists of four sequential stages: (1) the \textit{Solver} generates multiple candidate reasoning traces; (2) the \textit{Critic} refines each reasoning trace; (3) the \textit{Rewriter} synthesizes rewritten versions referencing all prior solutions; and (4) the \textit{Selector} ranks and outputs the final answer. For efficiency on our dataset, we reduce the number of reasoning iterations per role from five to three while maintaining identical temperature and decoding parameters. All stages are executed within a single inference session to ensure consistent context propagation across steps.

This adaptation preserves SciMaster's original tool-augmented reasoning design while ensuring compatibility with our scientific question-answering setting. The resulting baseline provides a robust reference point for evaluating the performance gains introduced by \agentname's reasoning-guided section ranking and iterative reading mechanisms.

\section{Experiments Compute Resources and Model Choices}

We conducted all experiments using API access to GPT-4o-20240806, GPT-4.1-20250414, and DeepSeek-R1-20250120, o3-20250416, o4-mini-20250416, gemini-2.5-pro-20250617, and Llama-3.1-70B-Instruct-Turbo-20250723 -- using their default settings.

Other experiments compute resources indicated in Table~\ref{tab:hardware}.

\begin{table}[h]
  \centering
  
  \small 
  \begin{tabular}{l l}
    \toprule
    \textbf{Component} & \textbf{Specification} \\
    \midrule
    GPU Model      & NVIDIA RTX 3090 \\
    GPU Memory     & 24 GB per GPU \\
    CPU Cores      & 22 \\
    System Memory  & 64 GB \\
    \bottomrule
  \end{tabular}
  \caption{Hardware Configuration Used for Training and Inference}
  \label{tab:hardware}
\end{table}

\newpage
\section{Detailed Impact of Mapping Model Across Domains}

To complement our primary analysis of mapping model robustness (Section Robustness of \agentname to Mapping Model and Input Variability), we present a comprehensive evaluation across all three scientific domains in \benchname: SERS in chemistry physics (Phys), infectious-disease modeling in public health (PH), and remote sensing in earth science (ES). 
This analysis investigates how different mapping models perform when paired with short-form answers generated by \agentname and vanilla RAG under different backbone LLMs.

Specifically, we follow a standard evaluation protocol: short-form answers are first generated by two methods using three backbone LLMs—GPT-4o, GPT-4.1, and DeepSeek-R1. These responses are then mapped to multiple-choice (MCQ) selections using three separate mapping models (GPT-4o, GPT-4.1, and DeepSeek-R1), resulting in 18 unique backbone-mapping combinations. Each configuration is evaluated using a consistent prompt template. Accuracy scores for each domain and setting are reported in Table~\ref{tab:model-performance_diff_J}.

Regardless of the mapping model used, \agentname consistently outperforms the RAG baseline across all domains and backbone configurations on average. The performance gains remain substantial, demonstrating the robustness of \agentname’s short-form answers across different downstream mapping strategies. 

Across all nine mapping–backbone pairs, \agentname consistently exceeds the vanilla RAG baseline: averaged over the three scientific domains, its accuracy gains are 4.8\%, 10.0\%, and 7.7\% when GPT-4o is the mapping model (for GPT-4o, GPT-4.1, and DeepSeek-R1 backbones, respectively); 8.0\%, 11.0\%, and 4.9\% when GPT-4.1 performs the mapping; and 5.6\%, 5.0\%, and 8.7\% when DeepSeek-R1 is used as the mapping model.

\begin{table*}[t!h]
  \centering

    \small
\setlength{\tabcolsep}{3.5pt}
\begin{tabular}{llcccccccccccc}
  \toprule
  \multicolumn{1}{l}{ } & \multicolumn{1}{l}{Backbone} & \multicolumn{4}{c}{GPT-4o} & \multicolumn{4}{c}{GPT-4.1} & \multicolumn{4}{c}{DeepSeek-R1} \\
  \cmidrule(r){3-6}\cmidrule(r){7-10}\cmidrule(r){11-14}
                                       Mapping & Method & Phys & PH & ES & Avg. & Phys & PH & ES & Avg. & Phys & PH & ES & Avg. \\
  \midrule
  \multirow{2}{*}{GPT-4o}      & \agentname & 73.8 & 52.7 & 60.0 & 62.2 & 76.9 & 63.6 & 68.3 & 69.6 & 73.8 & 67.3 & 63.3 & 68.1 \\
                               & RAG        & 63.1 & 50.9 & 58.3 & 57.4 & 66.2 & 52.7 & 60.0 & 59.6 & 63.1 & 58.2 & 60.0 & 60.4 \\
  \midrule
  \multirow{2}{*}{GPT-4.1}     & \agentname & 75.4 & 58.2 & 63.3 & 65.6 & 78.5 & 69.1 & 70.0 & 72.5 & 72.3 & 67.3 & 63.3 & 67.6 \\
                               & RAG        & 60.0 & 52.7 & 60.0 & 57.6 & 66.2 & 60.0 & 58.3 & 61.5 & 64.6 & 63.6 & 60.0 & 62.7 \\
  \midrule
  \multirow{2}{*}{DeepSeek-R1} & \agentname & 67.7 & 58.2 & 58.3 & 61.4 & 70.8 & 58.2 & 63.3 & 64.1 & 76.9 & 67.3 & 55.0 & 66.4 \\
                               & RAG        & 61.5 & 50.9 & 55.0 & 55.8 & 66.2 & 52.7 & 58.3 & 59.1 & 63.1 & 58.2 & 51.7 & 57.7 \\
  \bottomrule
\end{tabular}
\caption{Accuracy (in percentage) of our \agentname and the Vanilla RAG baseline across three scientific domains in \benchname—SERS in chemistry physics (Phys), infectious-disease modeling in public health (PH), and remote sensing in earth science (ES). 
Each row block corresponds to a mapping model (GPT-4o, GPT-4.1, or DeepSeek-R1), and each column block shows performance under a different backbone LLM.
Short-form answers from each backbone are mapped to MCQ choices using each mapping model.}
    \label{tab:model-performance_diff_J}
\end{table*}

\begin{table*}[t!]
\centering
\small
\begin{tabular}{lccc}
\hline
\textbf{Method} & \textbf{Top-1 Acc.} & \textbf{Top-2 Acc.} & \textbf{Top-3 Acc.} \\
\hline
Similarity-based (RAG) & 25.0 (14/56) & 32.0 (18/56) & 50.0 (28/56) \\
        Reasoning-based (\agentname) & 87.5(49/56) & 92.9(52/56) & 94.6(53/56)\\
\hline
\end{tabular}
\caption{Section title ranking accuracy (in percentage) on the physics dataset (Top-1, Top-2, and Top-3). Ground truth is based on expert-annotated sections.}
\label{tab:section-ranking-app}
\end{table*}

In addition, we revisited the results of human mapping annotation in collaboration with our physics domain expert. In a prior study, RAG achieved a score of 61 out of 65 when evaluated using GPT-4.1 as the mapping model. 
Upon closer inspection of the four mismatches, we found that in two cases the model's selections were arguably more accurate than those of the human annotator, indicating possible human error when the task is challenging (which causes potential ambiguities in answer interpretation).

Nevertheless, even with minor fluctuations in performance across different mapping models, we consistently observe a substantial performance gap between the best baseline and our approach. This further reinforces our conclusion that \agentname achieves state-of-the-art performance in \taskname.

\section{Similarity-Based Section Selection}

In addition to evaluating baseline approaches, we further investigate how to identify the section of a paper most relevant to a given research question by analyzing section titles. To this end, we conduct an additional experiment comparing two section selection strategies. The first is the reasoning-based ranking from the Section Ranking stage developed in \agentname. The second is a retrieval-based approach that selects the section whose title exhibits the highest cosine similarity with the research question, using a vanilla RAG framework.

This experiment focuses on the physics dataset, where domain experts have annotated the section titles from which correct answers are expected to originate. We first collect the section titles ranked by the reasoning-based method. For comparison, we replace this with the retrieval-based strategy described above, computing cosine similarity scores between each question and all section titles, and ranking them in descending order. Both methods are evaluated by comparing their top-3 ranked section titles against the expert-annotated ground truth. Results are reported in Table~\ref{tab:section-ranking-app}.

Note that in some cases where the correct answer is \textit{F (none of the above)}—indicating that the relevant content does not appear in the paper—we exclude those questions from the evaluation, as there is no valid section title to compare. As a result, the physics dataset contains 56 questions that are valid for Section Ranking comparison.

We observe that our method achieves a top-1 accuracy of 87.5\%, a top-2 accuracy of 92.9\%, and a top-3 accuracy of 94.6\%, in comparison to the RAG-based method, which yields an accuracy of 25.0\% at top-1, 32.0\% at top-2, and 50.0\% at top-3 on the physics dataset.

We also evaluate end-to-end task performance using the standard protocol. When replacing our reasoning-based section ranking with the RAG-based reordering strategy, the overall accuracy drops to 66.2\%, compared to 78.5\% achieved by \agentname using GPT-4.1 as both the backbone and mapping model. While this substitution leads to a performance decline, the decrease is not substantial. These results highlight the robustness of our iterative reading stage: even when section titles fail to generate an ideal ranking, the model continues to perform strongly by reasoning over multiple retrieved sections.

\end{document}

%% file: prompt.tex
\tcbset{
  promptstyle/.style={
    enhanced,
    breakable,
    colback=gray!3,
    colframe=gray!80!black,
    coltitle=white,
    colbacktitle=gray!70!black,
    fonttitle=\bfseries\small,
    boxrule=0.5pt,
    arc=2mm,
    outer arc=2mm,
    drop shadow southeast,
    left=6pt,
    right=6pt,
    top=6pt,
    bottom=6pt,
    before skip=10pt,
    after skip=10pt,
    sharp corners,
    verbatim=true    
  }
}

\section{Prompts}

Here is the list of prompts we used in our study.

\subsection{Section Ranking Prompts}\label{prompt:section-ranking-prompts}

\subsubsection{Hierarchy Preservation Prompt}\label{prompt:Hierarchy-Preservation-Prompt}
\begin{tcolorbox}[title=Prompt: Hierarchy Preservation, enhanced, breakable,
  colback=gray!3, colframe=gray!80!black,
  coltitle=white, colbacktitle=gray!70!black,
  fonttitle=\bfseries\small, boxrule=0.5pt,
  arc=2mm, outer arc=2mm, drop shadow southeast,
   sharp corners, before skip=2pt, after skip=4pt, parbox=false]
\begin{Verbatim}
hierarchy_preservation_prompt = """
You are an expert in scientific research and academic writing, proficient in analyzing the structural organization of research papers. Your task is to infer the hierarchical structure of a research paper based on a list of section titles provided in sequential order.

---

### **Task**
You are given a list of section titles extracted from a research paper, shown in the order they appear in the document:

{headings}

These titles are not annotated to indicate their level (e.g., section, subsection, sub-subsection). Your job is to infer the **tree structure** by determining which titles are:
- **Main sections**
- **Subsections** under a main section
- **Sub-subsections** under a subsection

---

### **Rules**
- Maintain the **original order** of the titles.
- Do **not** introduce any new titles or remove existing ones.
- Use the **first title** as the likely **main title** of the paper, unless it's clearly a preamble (e.g., "Abstract", "Introduction").
- Infer nesting levels based on typical academic paper structure and semantic cues in the titles.
- Structure the output using this format:
  - `"Section Title"` (main section)
  - `"Section Title - Subsection Title"` (subsection)
  - `"Section Title - Subsection Title - Sub-subsection Title"` (sub-subsection)

---

### **Example**

Given:

["Experiment setup", "sample preparation", "device setup", "SERS measurements", "Result and Discussion"]

Expected output:

["Experiment setup", 
 "Experiment setup - sample preparation", 
 "Experiment setup - device setup", 
 "Experiment setup - SERS measurements", 
 "Result and Discussion"]

---

### **Expected Output Format**
- Provide the inferred tree structure in a Python list format.
- Follow the form: `["Section", "Section - Subsection", "Section - Subsection - Sub-subsection", ...]`
- Output **only the list**. Do not include explanations, reasoning steps, or extra comments.

---

Let's think step by step.
"""
\end{Verbatim}
\end{tcolorbox}

\subsubsection{Reasoning-Based Ranking Prompt}\label{prompt:Reasoning-Based Ranking-Prompt}
\begin{tcolorbox}[title=Prompt: Reasoning-Based Ranking, enhanced, breakable,
  colback=gray!3, colframe=gray!80!black,
  coltitle=white, colbacktitle=gray!70!black,
  fonttitle=\bfseries\small, boxrule=0.5pt,
  arc=2mm, outer arc=2mm, drop shadow southeast,
  sharp corners, before skip=2pt, after skip=4pt, parbox=false]
\begin{Verbatim}
ranking_prompt = """
You are an AI research assistant with expertise in analyzing academic papers. Your task is to determine which sections of a paper are most likely to contain the answer to a given research question. The section titles are provided in a **tree structure**, where main sections and their subsections are denoted using a hyphen ("-"). Your goal is to rank these sections and subsections from **most relevant to least relevant** in relation to the research question.

---

**Paper Title**: {main_title_heading}

**Research Question**: {question}

**Section Titles**:
{formatted_headings}

---

### **Instructions**
- Analyze the research question and the structure of the section titles.
- Rank the section titles in order of likelihood to contain the answer — from most to least relevant.
- Subsections may be more specific than main sections; use their nesting to guide your ranking.
- For each section in the ranking, provide a **one-sentence explanation** for its placement.
- Maintain the original order of section titles when relevance is tied.
- Do **not** skip or omit any section titles from the list.
- Format the final ranking using only integer indices, enclosed in <<<#>>> brackets, based on the position of each title in the provided list (starting from 1).
- Do not repeat section titles or include any extra commentary after the final ranking.

---

### **Response Template**

Reasoning steps:

1. [Title 1] — [1-sentence justification]
2. [Title 2] — [1-sentence justification]
...

**Final ranking:** <<<2>>>, <<<1>>>, <<<3>>>, ... <<<n>>>

Let's think step by step.
"""
\end{Verbatim}
\end{tcolorbox}

\subsection{Iterative Reading Prompts}\label{prompt:Iterative Reading Prompts}
\subsubsection{Action Loop Prompt}\label{prompt:Action Loop Prompt}
\begin{tcolorbox}[title=Prompt: Action Loop, enhanced, breakable,
  colback=gray!3, colframe=gray!80!black,
  coltitle=white, colbacktitle=gray!70!black,
  fonttitle=\bfseries\small, boxrule=0.5pt,
  arc=2mm, outer arc=2mm, drop shadow southeast,
   sharp corners, before skip=2pt, after skip=4pt, parbox=false]
\begin{Verbatim}
main_prompt = """
You are an assistant designed to select the next Action based on the current observation. 

Each observation contains:
- The current chunk index, indicating which chunk of the document you are working on.
- The total number of available chunks.
- A record of past Actions taken in previous iterations.
- If the most recent Action was EVALUATE, the observation will also include the evaluation result.

---

To ensure accuracy, you must follow these instructions:

{main_prompt_instructions}

---

Descriptions of all allowed actions:

- GET_CHUNK: Retrieve the next Knowledge Chunk. Use this when no chunk is currently loaded or when advancing to the next chunk.
- GET_DETAIL: Extract relevant details with respect to the Research Question from the currently loaded Knowledge Chunk.
- EVALUATE: Determine whether the extracted details from the current Knowledge Chunk are sufficient to answer the Research Question.
- TERMINATE: End the process once you have gathered enough relevant information to confidently answer the Research Question. (TERMINATE is only valid if the most recent evaluation yielded a “YES” result.)

---

Here are some examples:

- Past Action Taken: GET_CHUNK. I have already obtained the chunk, so I will GET_DETAIL.
- Past Action Taken: GET_CHUNK, GET_DETAIL. then I will now EVALUATE the summary.
- Past Action Taken: GET_CHUNK, GET_DETAIL, EVALUATE. The result does not seem to be going well, so I will GET_CHUNK again.
- Past Action Taken: GET_CHUNK, GET_DETAIL, EVALUATE. The result seems to be going well, so I will TERMINATE the process.
- Past Action Taken: GET_CHUNK, GET_DETAIL, EVALUATE, ... ,GET_CHUNK, GET_DETAIL, EVALUATE. This is chunk 10 out of 10. I was supposed to get more chunks, but I have already retrieved the last one. No more knowledge chunks are available, so I will TERMINATE the process.

---

Observation:

You are currently at chunk {current_chunk_index} out of {total_chunks_len} chunks.

Past actions taken:
{past_action}

{evaluation_result}

---

Response Format:

Your final answer must follow the exact format below. Do not include any text outside this format.

Format:

Reasoning Steps: [In one sentence, explain why this Action is the best next step.]

Action: [Choose one of the following: GET_CHUNK, GET_DETAIL, EVALUATE, TERMINATE]
"""
\end{Verbatim}
\end{tcolorbox}

\subsubsection{Action Loop - Default Balanced Instruction}\label{prompt:Action Loop instruction Prompt}
For the instruction given in the loop above, depend on confidence level, different prompt was provided. This given Instruction subject to confidence level 2. This instruction is being used for default setting.
\begin{tcolorbox}[title=Prompt: Action Loop, enhanced, breakable,
  colback=gray!3, colframe=gray!80!black,
  coltitle=white, colbacktitle=gray!70!black,
  fonttitle=\bfseries\small, boxrule=0.5pt,
  arc=2mm, outer arc=2mm, drop shadow southeast,
   sharp corners, before skip=2pt, after skip=4pt, parbox=false]
\begin{Verbatim}
main_prompt_instructions_confidence_level_2 = """
1. You will be provided with a list of Knowledge Chunks, ordered by relevance to the research question. Earlier Knowledge Chunks are more likely to contain useful information.
2. Your task is to iteratively retrieve a Knowledge Chunk, extract relevant details, and evaluate whether you can answer the research question. Repeat this process until you can confidently terminate.
3. You will be given a list of predefined actions to select from.
4. In each iteration, you must select exactly one action based on the current observation.
5. The observation includes: the current Knowledge Chunk index, the total number of available Knowledge Chunks, and the list of past actions taken.
6. The available actions are: GET_CHUNK, GET_DETAIL, EVALUATE, and TERMINATE, as defined in the action list.
7. If the most recent action in the past actions taken was GET_CHUNK, your next action **must** be GET_DETAIL to extract information from the current Knowledge Chunk — do not perform this extraction yourself.
8. If the most recent action was GET_DETAIL, your next action **must** be EVALUATE to assess whether the gathered details are sufficient to answer the research question — do not perform the evaluation yourself.
9. You must select only one action at a time — never choose multiple actions in a single step.
10. Since Knowledge Chunks are ordered by relevance to the research question, earlier Knowledge Chunks are more likely to contain useful information. Use this ordering to guide your reading sequence.
11. You must TERMINATE the process once you have gathered enough information to confidently answer the research question.
12. Pay attention to the total number of Knowledge Chunks. If your most recent EVALUATE action was performed on the **last available Knowledge Chunk** and the result was insufficient, you must TERMINATE the process due to lack of additional information.
13. The expected workflow is: GET_CHUNK → GET_DETAIL → EVALUATE → TERMINATE (if evaluation result is sufficient). If not, continue to the next Knowledge Chunk. If you reach the final Knowledge Chunk and the evaluation is still insufficient, you must TERMINATE.
14. You may TERMINATE early if an EVALUATE action confirms that the gathered information is sufficient to answer the research question with high confidence.
15. After your reasoning, output your response in the specified format.
"""
\end{Verbatim}
\end{tcolorbox}

\subsubsection{Section Detail Extraction Prompt}\label{prompt:Section-Detail-Extraction-Prompt}
\begin{tcolorbox}[title=Prompt: Action: Section Detail Extraction,
  colback=gray!3, colframe=gray!80!black,
  coltitle=white, colbacktitle=gray!70!black,
  fonttitle=\bfseries\small, boxrule=0.5pt,
  arc=2mm, outer arc=2mm, drop shadow southeast,
   sharp corners, before skip=2pt, after skip=4pt, parbox=false]
\begin{Verbatim}
get_detail_prompt = """
You are a research assistant helping extract detailed information relevant to the given Research Question based on a Knowledge Chunk given.

---

Research Question: {question}

Knowledge Chunk:
{chunk}

---

Task:
- Extract all key points from the current Knowledge Chunk **only as they relate to the Research Question**.
- Include all relevant information such as scientific terms, numerical data, experimental results, measurements, statistical indicators, conclusions, and any comparative or causal statements that explicitly support or address the Research Question.
- Do not infer or assume missing content. If essential information is absent, clearly state what is missing—avoid speculation or completion.
- When a sentence directly answers or supports the Research Question, quote it **verbatim** from the **original chunk**.
- If there are multiple key points, present them as a structured list in bullet point format.
- If there are no key points relevant to the Research Question, clearly state: *"This chunk does not provide relevant information."*
- Perform this task step by step, ensuring that each extracted detail is justified and directly traceable to the original content.

---

Response Format:

Your final answer must follow the exact format below. Ensure strict adherence to this structure. Do not include additional explanations outside of this format.

Format:
Reasoning Steps: [reasoning step by step goes here]
Details: The Section_title is <Section_title>.[Details based on the Research Question and chunk go here. Include all relevant findings and quote supporting sentences, present them as a structured list.]
"""
\end{Verbatim}
\end{tcolorbox}

\subsubsection{Default Balanced Information Sufficiency Check Prompt}\label{prompt:Information-Sufficiency-Check-clvl2-Prompt}
The Information Sufficiency Check Prompt, confidence level 2. This prompt is being used for the default setting.
\begin{tcolorbox}[title=Prompt: Sufficiency Check  Confidence Level 2,
   enhanced jigsaw,
  breakable,
  width=\columnwidth,
  colback=gray!3,
  colframe=gray!80!black,
  coltitle=white,
  colbacktitle=gray!70!black,
  fonttitle=\bfseries\small,
  boxrule=0.5pt,
  arc=2mm,
  outer arc=2mm,
  drop shadow southeast,
  sharp corners,
  before skip=2pt,
  after skip=4pt]
\begin{Verbatim}
evaluation_prompt_confidence_level_2 = """
You are a research assistant tasked with evaluating whether the provided details are both sufficient and accurate to answer a given research question. Based solely on the current details, determine whether they contain all the necessary and correct information required to answer the research question. Your evaluation must include direct references to original sentences from the provided details to support your reasoning.

---

Research Question: {question}

Current Details:
{observation_stage}

---

Task:
- Assess whether the current details contain all essential and accurate information needed to answer the Research Question.
- Respond with "YES" **only if** the provided details fully and correctly address the research question with no missing elements or uncertainties.
  - In this case, provide a complete answer supported by quoted or clearly referenced content from the current Knowledge Chunk.
- If the information is incomplete, partially correct, or uncertain in any way, respond with "NO."
  - Clearly identify what specific information is missing or ambiguous.
  - Then provide the closest possible answer using only the available details.
  - If no relevant information is present at all, write: *"No information in given details."*

- After reasoning step by step, output both `Sufficiency` and `Detail_Answer`. Your `Detail_Answer` must include direct evidence from the knowledge chunk — avoid general summaries.

---

Response Format:

Your response must strictly follow the format below:

Sufficiency: [YES or NO]

Detail_Answer: The Section_title is <Section_title>. [Provide your best possible answer to the research question, using direct references from the details. Explain your reasoning step by step, ensuring each claim is supported by the provided content.]
"""
\end{Verbatim}
\end{tcolorbox}

\subsubsection{Detail Aggregation Final Answer Prompt}\label{prompt:Final-Answer-Prompts}
\begin{tcolorbox}[title=Prompt: Summary for Final Answer,
  enhanced jigsaw,
  breakable,
  width=\columnwidth,
  colback=gray!3,
  colframe=gray!80!black,
  coltitle=white,
  colbacktitle=gray!70!black,
  fonttitle=\bfseries\small,
  boxrule=0.5pt,
  arc=2mm,
  outer arc=2mm,
  drop shadow southeast,
  sharp corners,
  before skip=2pt,
  after skip=4pt]
\begin{Verbatim}
full_set_answer_prompt = """
You are a research assistant tasked with synthesizing a final answer to the research question based on the evaluations of multiple knowledge chunks provided below. Each evaluation entry includes:
- A sufficiency judgment (YES or NO),
- A detailed answer (Detail_Answer),
- Referenced original sentences, each tagged with its Chunk number and Section title.

Use the complete set of evaluation entries to construct a coherent, well-supported, and evidence-based final answer to the research question. You must include direct references to the **original sentences**, along with their corresponding **Chunk number** and **Section title**, to justify each claim you make.

---

Research Question: {question}

Evaluation Entries:
{observation_stage}

---

Task:
- Synthesize the provided evaluation entries to generate a comprehensive and conclusive answer to the Research Question.
- Your answer must be fully supported by specific content from the evaluation entries.
- When making a claim, **explicitly cite** the source using the format: `"Quoted sentence from the original text" (Chunk #, Section_title: <Section_title>)`.
- If multiple entries support the same point, cite each one.
- Do **not** introduce new information or inferred content that is not present in the evaluation entries.
- Avoid vague generalizations — every component of the answer must be evidence-backed.
- After reasoning step by step, output `Final_Answer` using the exact format below.

---

Response Format:

Final_Answer: [Provide your final answer with supporting evidence from the evaluation entries. For every claim, cite the original sentences along with the Chunk number and Section title they came from. Explain your reasoning step by step, covering all required aspects of the Research Question.]
"""
\end{Verbatim}
\end{tcolorbox}

\subsection{Evaluation Protocol Prompts}\label{subsec:Evaluation Protocol Prompt}
\subsubsection{Evaluation Protocol - Baseline Final Answer Prompt}\label{prompt:Baseline Final Answer Prompt}
\begin{tcolorbox}[title=Prompt: Final Answer - Choices Mapping,
   enhanced jigsaw,
  breakable,
  width=\columnwidth,
  colback=gray!3,
  colframe=gray!80!black,
  coltitle=white,
  colbacktitle=gray!70!black,
  fonttitle=\bfseries\small,
  boxrule=0.5pt,
  arc=2mm,
  outer arc=2mm,
  drop shadow southeast,
  sharp corners,
  before skip=2pt,
  after skip=4pt]
\begin{Verbatim}
full_set_answer_prompt = """
You are a research assistant tasked with synthesizing a final answer to the research question based on the evaluations of multiple knowledge chunks provided below. Each evaluation entry includes:
- A sufficiency judgment (YES or NO),
- A detailed answer (Detail_Answer),
- Referenced original sentences, each tagged with its Chunk number and Section title.

Use the complete set of evaluation entries to construct a coherent, well-supported, and evidence-based final answer to the research question. You must include direct references to the **original sentences**, along with their corresponding **Chunk number** and **Section title**, to justify each claim you make.

---

Research Question: {question}

Evaluation Entries:
{observation_stage}

---

Task:
- Synthesize the provided evaluation entries to generate a comprehensive and conclusive answer to the Research Question.
- Your answer must be fully supported by specific content from the evaluation entries.
- When making a claim, **explicitly cite** the source using the format: `"Quoted sentence from the original text" (Chunk #, Section_title: <Section_title>)`.
- If multiple entries support the same point, cite each one.
- Do **not** introduce new information or inferred content that is not present in the evaluation entries.
- Avoid vague generalizations — every component of the answer must be evidence-backed.
- After reasoning step by step, output `Final_Answer` using the exact format below.

---

Response Format:

Final_Answer: [Provide your final answer with supporting evidence from the evaluation entries. For every claim, cite the original sentences along with the Chunk number and Section title they came from. Explain your reasoning step by step, covering all required aspects of the Research Question.]
"""
\end{Verbatim}
\end{tcolorbox}

\subsubsection{Evaluation Protocol - Final Answer to MCQ Answer Mapping}\label{prompt: Final Answer to MCQ Answer Mapping}
\begin{tcolorbox}[title=Prompt: Final Answer - Final Answer to MCQ Answer Mapping,  enhanced jigsaw,
  breakable,
  width=\columnwidth,
  colback=gray!3,
  colframe=gray!80!black,
  coltitle=white,
  colbacktitle=gray!70!black,
  fonttitle=\bfseries\small,
  boxrule=0.5pt,
  arc=2mm,
  outer arc=2mm,
  drop shadow southeast,
  sharp corners,
  before skip=2pt,
  after skip=4pt]
\begin{Verbatim}
prompt = f"""
You are a research assistant. Your task is to map a short answer to a research question, derived from a scientific research paper, to the most appropriate option among the provided answer choices in a multiple-choice question (MCQ) format.
We are given the following research question:
{question}

To support your understanding of the research question, here is some relevant contextual information:
{additional_note}

Here are the extracted answers to the research question, taken from the research paper:
{relevant_chunks}

Here are the answer choices:
{paper_choice}

**Instructions:**
1. You must select **only one answer choice** from the list that is either correct or most relevant to the research question.
2. Provide your answer as a single letter (e.g., A, B, C, D, E, F) enclosed in the format <<<ANS>>>. Do not include any other text in your final output.
3. You may include step-by-step reasoning to justify your decision, but your final output must consist of only one answer choice in the required format.
4. This is a science and technology-related research question. For cases where the short answer involves numerical values:
    - You must map the short answer to the most accurate and relevant answer choice.
    - If the short answer and an answer choice represent numerical values, apply the following decimal precision alignment rules:
        - If the short answer and an answer choice have the **same number of decimal places**, compare them directly without rounding.
        - If they have **different numbers of decimal places**, round the value with **more digits** to match the **shorter decimal precision**, then compare.
        - After alignment, select the answer choice that **exactly matches** the short answer.
        - If no answer choice matches after applying these rules, select **“F. None of the above.”**
    - **Examples**:
        - Short answer = 1.1111, answer choices include 1.1 → round short answer to 1.1 → match → select 1.1
        - Short answer = 2.01, answer choices include 1.98 → both have two decimal places → no rounding → no match → select “F. None of the above”
5. Never provide multiple answers.
**Example of output format:**
*reasoning steps*
---
<<<C>>>
Let's think step by step.
"""
\end{Verbatim}
\end{tcolorbox}

\subsection{Confidence Level Ablation Study Prompts}\label{prompt:Confidence-Level}

\subsubsection{Confidence Level 1 Ablation Study - Conservative Information Sufficiency Check Prompt}\label{prompt:Confidence Level 1 Ablation Study - Conservative Sufficiency Check Prompt}
\begin{tcolorbox}[title=Prompt: Sufficiency Check Confidence Level 1,
  enhanced jigsaw,
  breakable,
  width=\columnwidth,
  colback=gray!3,
  colframe=gray!80!black,
  coltitle=white,
  colbacktitle=gray!70!black,
  fonttitle=\bfseries\small,
  boxrule=0.5pt,
  arc=2mm,
  outer arc=2mm,
  drop shadow southeast,
  sharp corners,
  before skip=2pt,
  after skip=4pt]
\begin{Verbatim}
evaluation_prompt_confidence_level_1 = """
You are a research assistant tasked with evaluating whether the provided details are both sufficient and accurate to answer a given research question. Based solely on the current details, determine whether they contain all the necessary and correct information required to answer the research question. You must act conservatively — your evaluation should follow the strictest standard: any ambiguity, missing element, or lack of clarity must result in a "NO".

---

Research Question: {question}

Current Details:
{observation_stage}

---

Task:
- Assess whether the extracted details contain all **necessary elements** to fully and unambiguously answer the Research Question.
- Only respond with "YES" if **all aspects** of the research question are addressed with precise and complete evidence from the provided knowledge chunk.
  - No assumptions, inferred logic, or external knowledge are allowed.
- If **any part** of the required information is missing, vague, incomplete, or unclear, respond with "NO".
  - Clearly identify what is missing or uncertain.
  - Then provide the closest possible answer that can be supported solely by the available content from the Knowledge Chunk.
  - If no relevant information is present at all, state: *"No information in given details."*

- After step-by-step reasoning, output both `Sufficiency` and `Detail_Answer`. Your `Detail_Answer` must include direct evidence from the knowledge chunk — avoid general summaries.

---

Response Format:

Your response must strictly follow the format below:

Sufficiency: [YES or NO]

Detail_Answer: The Section_title is <Section_title>. [Provide your best possible answer to the research question, using direct references from the details. Explain your reasoning step by step, addressing all required components.]
"""
\end{Verbatim}
\end{tcolorbox}

\subsubsection{Confidence Level 3 Ablation Study - Aggressive Information Sufficiency Check Prompt}\label{prompt:Confidence Level 3 Ablation Study - Aggressive Sufficiency Check Prompt}
\begin{tcolorbox}[title=Prompt: Sufficiency Check Confidence Level 3,
    enhanced jigsaw,
  breakable,
  width=\columnwidth,
  colback=gray!3,
  colframe=gray!80!black,
  coltitle=white,
  colbacktitle=gray!70!black,
  fonttitle=\bfseries\small,
  boxrule=0.5pt,
  arc=2mm,
  outer arc=2mm,
  drop shadow southeast,
  sharp corners,
  before skip=2pt,
  after skip=4pt]
\begin{Verbatim}
evaluation_prompt_confidence_level_3 = """
You are a research assistant tasked with evaluating whether the provided details are both sufficient and accurate to answer a given research question. You may act more aggressively and confidently. Based solely on the current details, determine whether they contain **enough relevant content to reasonably answer** the research question, even if some minor points are not fully explicit.

---

Research Question: {question}

Current Details:
{observation_stage}

---

Task:
- Assess whether the current details provide **most or all of the key information** needed to reasonably answer the Research Question.
- Respond with "YES" if the answer can be supported using the provided information, even if a few supporting details are missing, implicit, or only partially clear.
  - If confident, provide a full answer backed by the best available references from the Knowledge Chunk.
- Only respond with "NO" if **critical** information is missing or the answer would be too speculative.
  - Clearly explain what is missing or unclear.
  - Then provide the closest possible answer based only on the available content.
  - If no relevant information is present at all, state: *"No information in given details."*

- After step-by-step reasoning, output both `Sufficiency` and `Detail_Answer`. Your `Detail_Answer` must include direct evidence from the knowledge chunk — avoid general summaries.

---

Response Format:

Your response must strictly follow the format below:

Sufficiency: [YES or NO]

Detail_Answer: The Section_title is <Section_title>. [Provide your best possible answer to the research question, using direct references from the details. Explain your reasoning step by step, addressing all required components.]
"""
\end{Verbatim}
\end{tcolorbox}

\subsubsection{Confidence Level 1 Ablation Study - Action Loop  Conservative Instruction}\label{prompt:Action Loop Instruction clvl1}
\begin{tcolorbox}[title=Prompt: Confidence Level Case Study Prompts 1,
    enhanced jigsaw,
  breakable,
  width=\columnwidth,
  colback=gray!3,
  colframe=gray!80!black,
  coltitle=white,
  colbacktitle=gray!70!black,
  fonttitle=\bfseries\small,
  boxrule=0.5pt,
  arc=2mm,
  outer arc=2mm,
  drop shadow southeast,
  sharp corners,
  before skip=2pt,
  after skip=4pt]
\begin{Verbatim}
main_prompt_instructions_confidence_level_1 = """
1. You will be provided with a list of Knowledge Chunks, ordered by relevance to the research question. Earlier Knowledge Chunks are more likely to contain useful information.
2. Your task is to iteratively retrieve a Knowledge Chunk, extract relevant details, and evaluate whether you can answer the research question. Repeat this process until you have evaluated all available Knowledge Chunks or gathered enough information to confidently terminate.
3. You will be given a list of predefined actions to select from.
4. In each iteration, you must select exactly one action based on the current observation.
5. The observation includes: the current Knowledge Chunk index, the total number of available Knowledge Chunks, and the list of past actions taken.
6. The available actions are: GET_CHUNK, GET_DETAIL, EVALUATE, and TERMINATE, as defined in the action list.
7. If the most recent action in the past actions taken was GET_CHUNK, your next action **must** be GET_DETAIL to extract information from the current Knowledge Chunk — do not perform this extraction yourself.
8. If the most recent action was GET_DETAIL, your next action **must** be EVALUATE to assess whether the gathered details are sufficient to answer the research question — do not perform the evaluation yourself.
9. You must select only one action at a time — never choose multiple actions in a single step.
10. Since Knowledge Chunks are ordered by relevance to the research question, earlier Knowledge Chunks are more likely to contain useful information. Use this ordering to guide your reading sequence.
11. You must TERMINATE the process only after either (a) all available Knowledge Chunks have been evaluated, or (b) you are confident that the research question can be answered based on the most recent evaluation.
12. Pay attention to the total number of Knowledge Chunks. If your most recent EVALUATE action was performed on the **last available Knowledge Chunk**, you must TERMINATE the process regardless of the outcome.
13. The expected workflow is: GET_CHUNK → GET_DETAIL → EVALUATE → TERMINATE (if evaluation result is sufficient). If not, continue to the next chunk. If you reach the final chunk and the evaluation is still insufficient, you must TERMINATE.
14. Do not TERMINATE early based solely on assumptions about relevance. Always continue until a conclusive evaluation result is available or all chunks are exhausted.
15. After your reasoning, output your response in the specified format.
"""
\end{Verbatim}
\end{tcolorbox}

\subsubsection{Confidence Level 3 Ablation Study - Action Loop Aggressive Instruction}\label{prompt:Action Loop Instruction clvl3}
\begin{tcolorbox}[title=Prompt: Confidence Level Case Study Prompts 3,
   enhanced jigsaw,
  breakable,
  width=\columnwidth,
  colback=gray!3,
  colframe=gray!80!black,
  coltitle=white,
  colbacktitle=gray!70!black,
  fonttitle=\bfseries\small,
  boxrule=0.5pt,
  arc=2mm,
  outer arc=2mm,
  drop shadow southeast,
  sharp corners,
  before skip=2pt,
  after skip=4pt]
\begin{Verbatim}
main_prompt_instructions_confidence_level_3 = """
1. You will be provided with a list of knowledge chunks, ordered by relevance to the research question. Earlier chunks are more likely to contain useful information.
2. Your task is to iteratively retrieve a knowledge chunk, extract relevant details, and evaluate whether you can answer the research question. Repeat this process until you can confidently terminate.
3. You will be given a list of predefined actions to select from.
4. In each iteration, you must select exactly one action based on the current observation.
5. The observation includes: the current chunk index, the total number of available chunks, and the list of past actions taken.
6. The available actions are: GET_CHUNK, GET_DETAIL, EVALUATE, and TERMINATE, as defined in the action list.
7. If the most recent action in the past actions taken was GET_CHUNK, your next action **must** be GET_DETAIL to extract information from the current knowledge chunk — do not perform this extraction yourself.
8. If the most recent action was GET_DETAIL, your next action **must** be EVALUATE to assess whether the gathered details are sufficient to answer the research question — do not perform the evaluation yourself.
9. You must select only one action at a time — never choose multiple actions in a single step.
10. Since knowledge chunks are ordered by relevance to the research question, earlier chunks are more likely to contain useful information. Use this ordering to guide your reading sequence.
11. You may TERMINATE the process as soon as you believe that additional knowledge chunks are unlikely to provide significantly better information, even if you are not fully confident.
12. If your most recent EVALUATE action was performed on the **last available chunk**, you must TERMINATE the process regardless of the outcome.
13. The expected workflow is: GET_CHUNK → GET_DETAIL → EVALUATE → TERMINATE. You may also TERMINATE early based on the assumption that remaining chunks have diminishing relevance.
14. Terminate aggressively if your evaluation suggests NO from continuing, even if high confidence has not yet been achieved.
15. After your reasoning, output your response in the specified format.
"""
\end{Verbatim}
\end{tcolorbox}

\subsection{Contextual RAG Prompt}\label{prompt:contexual RAG}
\begin{tcolorbox}[title=Prompt: Contextual RAG Prompt,
   enhanced jigsaw,
  breakable,
  width=\columnwidth,
  colback=gray!3,
  colframe=gray!80!black,
  coltitle=white,
  colbacktitle=gray!70!black,
  fonttitle=\bfseries\small,
  boxrule=0.5pt,
  arc=2mm,
  outer arc=2mm,
  drop shadow southeast,
  sharp corners,
  before skip=2pt,
  after skip=4pt]
\begin{Verbatim}
prompt = """
You are an expert technical writer specialised in contextual retrieval.

<document>
{whole_document}
</document>

Here is the chunk we want to situate within the whole document
<chunk>
{chunk}
</chunk>

Please give a short succinct context to situate this chunk within
the overall document for the purposes of improving search retrieval
of the chunk. Answer only with the succinct context and nothing else.
"""
\end{Verbatim}
\end{tcolorbox}